\documentclass[prd,reprint,nofootinbib,superscriptaddress,preprintnumbers,showpacs]{revtex4-1}
\usepackage{amsfonts}
\usepackage{amssymb}
\usepackage{amsmath}
\usepackage{graphicx,color}
\usepackage{dcolumn}
\usepackage{epsfig}
\usepackage{bm}
\usepackage{bbm}
\usepackage{ulem}
\usepackage{hyperref}
\usepackage{lineno}
\usepackage{multirow}
\usepackage{slashed}
\usepackage{makecell}
\setcellgapes{1pt}
\makegapedcells
\usepackage{bigdelim}
\makeatletter
\@ifundefined{footref}{
  \newcommand{\footref}[1]{\textsuperscript{\ref{#1}}}
}{}
\makeatother
\usepackage{color}
\newcommand{\red}[1]{\textcolor{black}{#1}}

\newcommand{\ket}[1]{|{#1}\rangle}

\def\gtap{\ \raise.3ex\hbox{$>$\kern-.75em\lower1ex\hbox{$\sim$}}\ }
\def\ltap{\ \raise.3ex\hbox{$<$\kern-.75em\lower1ex\hbox{$\sim$}}\ }
\begin{document}
\title{
Two nearby states in the $X(3872)$ region:\\
Resolving the radiative-decay ratio tension with $\eta_{c2}$}
\author{Satoshi X. Nakamura}
\email{satoshi@sdu.edu.cn}
\affiliation{
Institute of Frontier and Interdisciplinary Science, Shandong
University, Qingdao, Shandong 266237, China
}
 
\begin{abstract}
Recently, LHCb reported the radiative-decay ratio
${\cal R}^{\psi\gamma}\equiv 
{\cal B}[X(3872)\to \psi'\gamma]/{\cal B}[X(3872)\to J/\psi\gamma]=1.67\pm 0.25$
extracted from $B^+\to K^+(J/\psi\gamma, \psi'\gamma)$.
This result
differs markedly ($\sim4.6\sigma$) from the BESIII value obtained from
$e^+e^-\to \gamma(J/\psi\gamma, \psi'\gamma)$,
${\cal R}^{\psi\gamma}=-0.04\pm 0.28$.
Such a significant tension suggests 
that more than one state in the $X(3872)$ region contributes to the processes.
We therefore propose a two-state scenario: a shallow $D^{*0}\bar{D}^0$ bound state
with $J^{PC}=1^{++}$ and a $2^{-+}$ charmonium candidate, $\eta_{c2}$,
 slightly above the $D^{*0}\bar{D}^0$ threshold.
We show that this hypothesis consistently describes these ratios
along with other branching fractions and lineshapes across
multiple processes.
By contrast, fits without the $\eta_{c2}$ component fail to reproduce
 the radiative-ratio data.
We also predict helicity-angle distributions that motivate
the future experiments to test
the two-state hypothesis and search for the so-far missing $\eta_{c2}$.
\end{abstract}

\maketitle

\section{Introduction}

The 2003 discovery of $X(3872)$~\cite{x3872_belle_1}
ignited a modern era of hadron spectroscopy~\cite{review_chen,review_hosaka,review_lebed,review_esposito,review_ali,review_guo,review_olsen,review_Brambilla,review_rome}
by highlighting limitations of the traditional quark model~\cite{GI,BG}.\footnote{
We follow the particle naming convention of the Particle Data Group
(PDG)~\cite{pdg}.
We also denote $\psi(2S)$, $\psi(4230)$, and $\chi_{c1}(3872)$ by 
$\psi'$, $Y(4230)$, and $X(3872)$, respectively.
The notation $\{D^{*}\bar{D}\}$ indicates the positive $C$-parity state
such as $(D^{*0}\bar{D}^0+\bar{D}^{*0}{D}^0)/\sqrt{2}$ in our
convention.} 
Lying essentially at the $D^{*0}\bar{D}^0$ threshold,
$X(3872)$ exhibits a dual character: its mass and the narrow width favor a loosely bound
hadronic molecule~\cite{swanson2004,BKN2004,Lee2009,Guo2013},
while its prompt production rates point to a more compact
component~\cite{suzuki2005,Kalashnikova2005,Bignamini2009,cms_x3872_pp,Meng2017,lhcb_mult2021,Esposito2021}.
Lattice QCD also found an $s$-wave $D^{*0}\bar{D}^0$ bound state
consistent with $X(3872)$~\cite{Prelovsek2013,Padmanath2015}.

Here we point out that existing data actually strain the assumption that
a single state dominates the $X(3872)$ region.
To quantify this, we define the ratio
\begin{eqnarray}
 \label{eq:RB}
{\cal R}_{B}(f,f') = {
\Gamma[B^+ \to K^+ X(3872); X(3872)\to f]
\over 
\Gamma[B^+ \to K^+ X(3872); X(3872)\to f']
},
\end{eqnarray}
with $f^{(\prime)} = J/\psi\pi^+\pi^-, J/\psi\omega, D^{*0}\bar{D}^0+{\rm c.c.}, J/\psi\gamma$, 
and $\psi' \gamma$; 
$\Gamma[B^+ \to K^+ X; X\to f^{(\prime)}]$
is an $X(3872)$ contribution to
a $B^+\to K^+f^{(\prime)}$ partial decay width,
obtained through experimental analysis.
Similarly, ${\cal R}_{e^+e^-}$ is defined 
by replacements $\Gamma\to\sigma$ (cross section),
$B^{+}\to e^+e^-$ and $K^+\to\gamma$ in Eq.~(\ref{eq:RB}).

We expect
${\cal R}_B(f,f')\sim {\cal R}_{e^+e^-}(f,f')\sim {\cal B}[X(3872)\to f]/{\cal B}[X(3872)\to f']$, 
assuming that the experimental $\Gamma[B^+ \to K^+ X; X\to f^{(\prime)}]$ contain
sufficiently isolated $X(3872)$ contributions.
However, previous experiments found
${\cal R}_B(J/\psi\gamma,J/\psi\pi^+\pi^-)=0.22\pm 0.07$\footnote{Statistical and systematic
uncertainties are summed in quadrature throughout this paper.}~\cite{x3872_belle_jpsi-gamma,x3872_lhcb_jpsi-rho}
and
${\cal R}_{e^+e^-}(J/\psi\gamma,J/\psi\pi^+\pi^-)=0.79\pm 0.28$~\cite{x3872_bes3_DstarDbar},
giving a difference of $\sim$2$\sigma$.
Even more strikingly,
${\cal R}_{x}^{\psi\gamma}\equiv {\cal R}_x(\psi'\gamma,J/\psi\gamma)$ 
with $x=B$ or $e^+e^-$ were reported to be
${\cal R}_{B}^{\psi\gamma}=1.67\pm 0.25$~\cite{x3872_lhcb_jpsi-gamma2} and
${\cal R}_{e^+e^-}^{\psi\gamma}=-0.04\pm 0.28$~\cite{x3872_bes3_DstarDbar},
giving a difference of $\sim$4.6$\sigma$.

Reconciling these results with a single $X(3872)$ state would require an
unexpectedly strong process dependence of ${\cal R}_{x}^{\psi\gamma}$,
for example from residual backgrounds, fit-model assumptions, or
underestimated systematic uncertainties.
Rather, the puzzling ${\cal R}_{x}^{\psi\gamma}$
can be naturally understood by considering
two distinct states $X_A$ and $X_B$, and supposing that:
(i)~$\psi'\gamma$ ($J/\psi\gamma$) events arise predominantly from $X_A$ ($X_B$) decays;
(ii) the production of $X_A$ relative to $X_B$ is substantially enhanced
in $B^+$ decays compared with $e^+e^-$ annihilations.

\begin{figure*}[t]
\begin{center}
\includegraphics[width=1\textwidth]{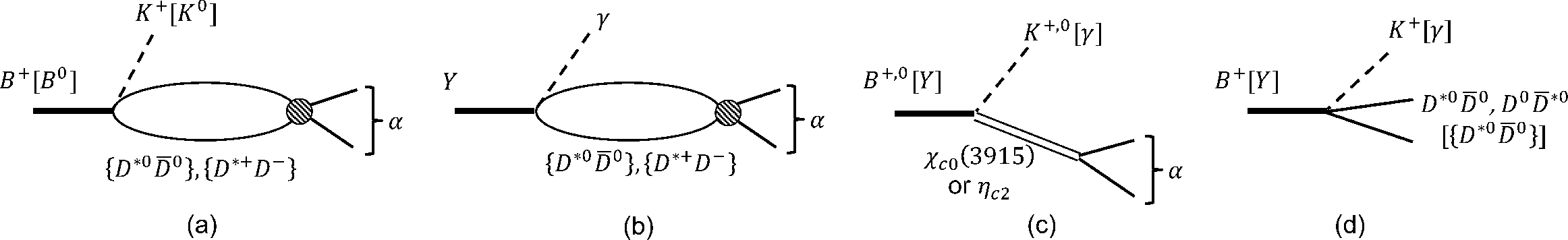}
\end{center}
 \caption{
$B^{+,0}$ and $Y(4230)$ decay mechanisms. 
Panels (a) and (b) show $B^{+,0}$ and $Y(4230)$ decays into open-charm
 intermediate states, respectively, 
followed by coupled-channel rescattering. 
The shaded blobs are the rescattering amplitude that generates the $X(3872)$ pole.
Panel (c) shows the $\eta_{c2}$- and $\chi_{c0}(3915)$-excitation mechanisms. 
Panel (d) shows the tree-level mechanisms. 
Final states $\alpha$ include $J/\psi\rho^0$, $J/\psi\omega$,
 $\{D^{*0}\bar{D}^{0}\}$, $J/\psi\gamma$, and $\psi'\gamma$. 
 }
\label{fig:diag}
\end{figure*}

We next identify the two inferred contributions.
From analyses of $B^+ \to K^+(J/\psi \pi^+ \pi^-)$, the $X(3872)$
properties are well constrained, with
$m_{X(3872)}=3871.64\pm 0.06$~MeV~\cite{x3872_lhcb_jpsi-rho}
\footnote{
We denote a particle $x$'s mass, momentum, energy, width, and spin state
in the total center-of-mass (CM) frame
by $m_x$, $\bm{p}_x$, $E_x$, $\Gamma_x$, and $s_x^z$,
respectively;
$E_x=\sqrt{m_x^2+p_x^2}$ with
$p_x=|\bm{p}_x|$.
The mass and width values are from the PDG~\cite{pdg}.
See Appendix~\ref{app1d}
 for the energy dependence of 
$\Gamma_x$.
}
and $J^{PC}=1^{++}$~\cite{lhcb2013}.
It is therefore natural to identify $X_A$ with the established
$X(3872)$, interpreted as a shallow $D^{*0}\bar{D}^0$ bound state (with
a small $c\bar{c}$ admixture) that
accounts for the observed isospin-violating decay
strength~\cite{HD2013,Guo2014} and 
accommodates a wide range of relative branching ratios to 
$\psi'\gamma$ and $J/\psi\gamma$~\cite{Ortega2013,Guo2015,Cincioglu2016,Takeuchi2017}.

Regarding $X_B$, 
previous analyses hinted at
a nearby $2^{-+}$ charmonium ($\eta_{c2}$)
decaying into $J/\psi \omega$~\cite{x3872_babar_omega} and
$D^{*0}\bar{D}^0$~\cite{x3872_babar_DstarD}.
An $\eta_{c2}$ in this mass region is also expected in quark models~\cite{GI,BG,Fulcher} and lattice QCD~\cite{lqcd_jlab,lqcd_eta_c2}, but has not yet been observed~\cite{belle_eta_c2,belle_eta_c2_2}.
Moreover, its radiative decay is expected to favor $J/\psi\gamma$ over $\psi'\gamma$~\cite{eta_c2_radiative_decay1,eta_c2_radiative_decay2,eta_c2_radiative_decay3}.
We therefore identify $X_B$ with an $\eta_{c2}$ lying slightly above the
$D^{*0}\bar{D}^0$ threshold.
An isospin-violating $\eta_{c2}$ decay, expected to be small, would not
affect the $J^{PC}$ determination using the $J/\psi\pi^+\pi^-$ data~\cite{lhcb2013}.
This two-state picture is also consistent with the observation that
the $X(3872)$ mass extracted from
$J/\psi\,\omega$~\cite{x3872_babar_omega,x3872_bes3_jpsi-omega} and $D^{*0}\bar{D}^0$~\cite{x3872_babar_DstarD,x3872_belle_DstarD2}
data tends to be 2--3~MeV higher than that from $J/\psi\pi^+\pi^-$~\cite{x3872_lhcb_jpsi-rho,x3872_bes3_jpsipipi}.
Although several issues in assigning $2^{-+}$ to $X(3872)$ have been critically
examined~\cite{eta_c2_burns,eta_c2_KN,eta_c2_Ke},
most of them do not apply to the two-state hypothesis.

In this work, we develop a two-state model that consistently describes a broad set of $X(3872)$ data and highlights clear advantages over a single-state description.
We also predict helicity-angle distributions with which the future experiments
can test the two-state hypothesis and search for
the missing $\eta_{c2}$.

\section{Method}

We analyze $B^{+(0)}\to K^{+(0)}f$ and $e^+e^-\to\gamma f$ data using the decay mechanisms summarized in Fig.~\ref{fig:diag}.
For $e^+e^-\to\gamma f$, we work at $\sqrt{s}=m_{Y(4230)}$ and model the production as
$e^+e^-\to Y(4230)\to\gamma f$, following Refs.~\cite{Guo2013,x3872_bes3_jpsi-omega}.
In Figs.~\ref{fig:diag}(a,b), $X(3872)$ emerges as a pole of a
$J^{PC}=1^{++}$ coupled-channel system dominated by the $\{D^{*}\bar{D}\}$ channels.
Any compact $c\bar{c}$ component~\cite{chic1_2p_cc,chic1_2p_lqcd} is
treated effectively through short-range contact interactions.
We additionally include an explicit $\eta_{c2}$-excitation mechanism
[Fig.~\ref{fig:diag}(c)] to address the radiative-ratio tension.
To describe peaks at $M_{J/\psi\omega}\sim 3.92$~GeV~\cite{x3872_babar_omega,x3872_bes3_jpsi-omega}, 
we follow the BABAR analysis~\cite{x3915_babar} and 
include the $\chi_{c0}(3915)$-excitation mechanism [Fig.~\ref{fig:diag}(c)].
An alternative $2^+$ state~\cite{x3915_ZXZ}
would not significantly change our result since 
$1^+$ $X(3872)$ interferes with neither 
$0^+$ nor $2^+$ states in the $M_{J/\psi\omega}$ distribution.

For the $X(3872)$ production mechanism in $B$-decay [Fig.~\ref{fig:diag}(a)], 
the initial weak vertex $B\to D^*\bar{D}K$ is parameterized as
\begin{eqnarray}
\label{eq:b_weak}
v^{B}_{D^*\bar{D}K} = 
c^{B}_{D^*\bar{D}K}\,
 f_{D^*\bar{D}}^{0}
 F_{K B}^{1}\, 
\bm{p}_{K}\cdot\bm{\epsilon}_{D^*} ,
\end{eqnarray}
with coupling constant $c^{B}_{D^*\bar{D}K}$.
The $B\to D\bar{D}^* K$ vertex 
with $c^{B}_{D\bar{D}^*K}$ is similar.
The $B^+$ decays include
color-favored couplings~\cite{x3872_color_favor}
$c^{B^+}_{D^{*0}\bar{D}^0K^+}$ and $c^{B^+}_{D^0\bar{D}^{*0}K^+}$ 
($c^{B^+}_{D^{*0}\bar{D}^0K^+}\ne c^{B^+}_{D^0\bar{D}^{*0}K^+}$)
and color-suppressed couplings
$c^{B^+}_{D^{*+}D^-K^+}$ and $c^{B^+}_{D^+D^{*-}K^+}$ 
($c^{B^+}_{D^{*+}D^-K^+}=c^{B^+}_{D^+D^{*-}K^+}$).
Regarding the weak $B^0$-decay couplings, 
we use relations:
$c^{B^0}_{D^{*+}D^-K^0}=c^{B^+}_{D^{*0}\bar{D}^0K^+}$, 
$c^{B^0}_{D^+D^{*-}K^0}=c^{B^+}_{D^0\bar{D}^{*0}K^+}$,
and 
$c^{B^0}_{D^{*0}\bar{D}^0K^0}=c^{B^0}_{D^0\bar{D}^{*0}K^0}=c^{B^+}_{D^{*+}D^-K^+}$.
The vertices in Eq.~(\ref{eq:b_weak}) include dipole form factors,
$f_{ij}^{L}(q_{ij},\Lambda_B)$ and $F_{kl}^{L'}(p_k,\Lambda)$,
\begin{eqnarray}
\label{eq:ff_weak}
 f_{ij}^{L}\! &=&\!\!
 \frac{(1+q_{ij}^2/\Lambda^2_B)^{-2-{L\over 2}}}{\sqrt{4E_i E_j}} ,
F_{kl}^{L'}\! =\!
 \frac{(1+p_k^2/\Lambda^{2})^{-2-{L'\over 2}}}{\sqrt{4E_k E_l}} ,
\end{eqnarray}
where $q_{ij}$ (${p}_{k}$) and $L$ ($L'$)
are the momentum and orbital angular momentum of $i$ ($k$) in the $ij$
(total) CM frame, respectively.
The cutoff $\Lambda_B$, chosen to be independent of $ij$,
is fitted to data, while 
$\Lambda=1$~GeV is fixed.

\begin{figure*}
\includegraphics[width=1.0\textwidth]{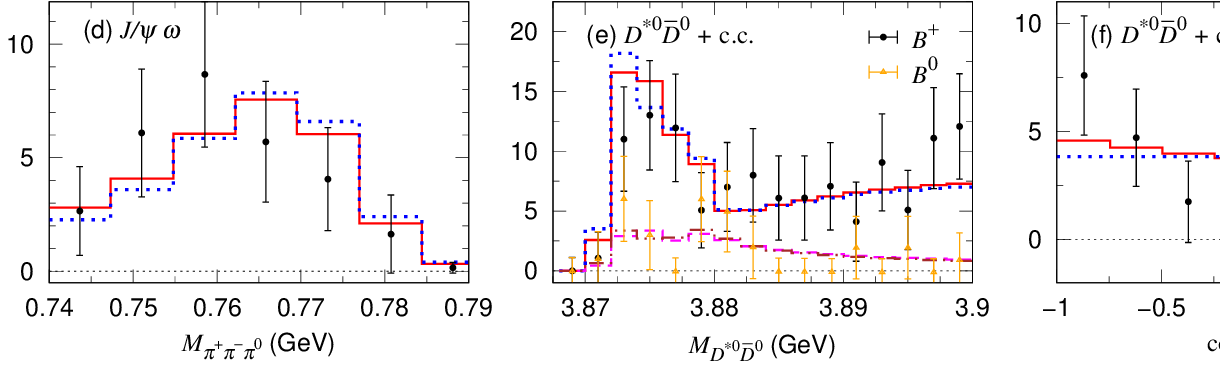}
 \caption{\label{fig:comp-data}
(a)-(e) Invariant mass $M_f$
and (f) helicity angle $\theta_{X(3872)}$
 distributions
 for $B\to Kf$ where final states $f$ are indicated in each panel;
$B^+\to K^+f$ for (a)-(e), $B^0\to K^0f$ for (e), and 
an efficiency-weighted sum of $B^+\to K^+f$ and $B^0\to K^0f$ in (f).
The units are events/bin.
Histograms are obtained 
by smearing the corresponding theoretical curves 
with experimental resolutions and then averaging within each bin.
In (c), the lower histograms and the data [orange triangle] are obtained
by cutting off contributions from $M_{\pi^+\pi^-\pi^0}<0.7695$~GeV.
The legend in (a) applies to all panels except (b), which has its own legend.
The data are from Ref.~\cite{x3872_lhcb_lineshape} in (a);
\cite{x3872_lhcb_pipi} in (b);
\cite{x3872_babar_omega} in (c) and (d);
\cite{x3872_belle_DstarD2} ($D^{*0}\to D^0\pi^0$) in (e);
\cite{x3872_babar_DstarD} in (f).
The uncorrected data points [orange triangle] in (c) are scaled to 
be comparable with the corrected data [black circles].
}
 \end{figure*}

The weak decays [Eq.~(\ref{eq:b_weak})]
are followed by coupled-channel rescatterings among the
$\{D^{*0}\bar{D}^0\}-\{D^{*+}D^-\}-J/\psi\omega-J/\psi\rho-J/\psi\gamma-\psi'\gamma$
channels.
We model the coupled-channel dynamics with separable $s$-wave potentials,
\begin{eqnarray}
v_{\beta,\alpha}(q',q) &=& h_{\beta,\alpha}\,f^0_\beta(q',\Lambda_v) \,  f^0_\alpha(q,\Lambda_v) ,
\label{eq:cont-ptl}
\end{eqnarray}
where $\alpha$ and $\beta$ label the channels, $h_{\beta,\alpha}$ are
coupling constants, and $\Lambda_v$ is a cutoff.
Motivated by the OZI rule, we set $h_{\beta,\alpha}=0$ if both $\alpha$
and $\beta$ are in hidden-charm channels.
We treat electromagnetic interactions to first order.
The resulting $B^+\to K^+\alpha$ decay amplitude is
\begin{widetext}
\begin{eqnarray}
A^{(a)}_{B^+\to K^+\alpha}
 &=&
\bm{p}_{K^+}\cdot\bm{\epsilon}_\alpha
\sum_{\beta'}
 f_{\alpha}^{0}(q_\alpha,\Lambda_v)\,
G_{\alpha,\beta'}(p_{K^+},E)
{h_{\beta',\beta}\over \sqrt{2}}
\left(
\sum_x^{D^{*}\bar{D},D\bar{D}^{*}}
c^{B^+}_{x K^+} 
 \sigma^{\Lambda_v\Lambda_B}_{\beta}(p_{K^+},E)\right)
 F_{K^+ B^+}^{1}(p_{K^+},\Lambda)\ , 
\label{eq:B_decay}
\end{eqnarray}
\end{widetext}
with $E=m_{B^+}-E_{K^+}$;
$\beta=\{D^{*0}\bar{D}^0\}$ for 
$x=D^{*0}\bar{D}^0,D^{0}\bar{D}^{*0}$ and 
$\beta=\{D^{*+}D^-\}$ for 
$x=D^{*+}D^{-}, D^{+}D^{*-}$;
$\bm{\epsilon}_\alpha$ is the total spin polarization of $\alpha$.
The $1/\sqrt{2}$ factor is due to projecting out
the $C=+1$ component of $x$.
We introduced
\begin{eqnarray}
[G^{-1}(p,E)]_{\beta\alpha}&=&\delta_{\beta\alpha} - h_{\beta,\alpha}
 \sigma^{\Lambda_v\Lambda_v}_{\alpha}(E) , 
\label{eq:green}
\end{eqnarray}
and~\cite{pipin}
\begin{eqnarray}
\sigma^{\Lambda_1\Lambda_2}_{\alpha}(p,E) &=&
\int\! dq q^2 
{M_{\alpha}(q)\over \sqrt{M^2_{\alpha}(q) + p^2}}
\nonumber\\
&&\times
{ 
f^0_\alpha(q,\Lambda_1) f^0_\alpha(q,\Lambda_2)
\over E - \sqrt{M^2_{\alpha}(q) + p^2} + i{\Gamma_{\alpha_1}+\Gamma_{\alpha_2}\over 2} 
} ,
\label{eq:sigma}
\end{eqnarray}
with $M_{\alpha}(q) = E_{\alpha_1}(q) + E_{\alpha_2}(q)$;
$\alpha_1$ and $\alpha_2$ are mesons forming $\alpha$.
The Green function $G_{\beta\alpha}$ of Eq.~(\ref{eq:green})
violates the isospin due to the mass difference
between $\{D^{*0}\bar{D}^0\}$ and $\{D^{*+}D^-\}$ channels.
A corresponding $B^0\to K^0\alpha$ amplitude follows from
Eq.~(\ref{eq:B_decay}) by replacing $B^+\to B^0$ and $K^+\to K^0$.

The $Y\to\gamma\alpha$ amplitude for Fig.~\ref{fig:diag}(b) is obtained analogously.
The $\eta_{c2}$ and $\chi_{c0}(3915)$ excitation contributions to $B$
and $Y$ decays [Fig.~\ref{fig:diag}(c)] are modeled with Breit--Wigner
forms.
We fit $m_{\eta_{c2}}$ to the data while taking a non-open-charm (nOC) partial width
as $\Gamma^{\rm nOC}_{\eta_{c2}}=0.5$~MeV~\cite{eta_c2_radiative_decay3}.
The $\chi_{c0}(3915)$ parameters are taken from Ref.~\cite{x3872_babar_omega}.
For tree-level $B^+$ decay contributions [Fig.~\ref{fig:diag}(d)],
we use Eq.~(\ref{eq:b_weak}) and its $D\bar{D}^*$ counterpart;
a tree-level $Y$ decay amplitude is similar.
LHCb observed an $\omega-\rho$ mixing effect
in the $M_{\pi^+\pi^-}$ distribution
of $B^+\to K^+ J/\psi\pi^+\pi^-$~\cite{x3872_lhcb_pipi}.
We add this mixing effect to the amplitudes discussed above. 
We also incorporate $\rho^0\to\pi^+\pi^-$ and $\omega\to\pi^+\pi^-\pi^0$ decays when
$\alpha=J/\psi\rho^0$ and $J/\psi\omega$, respectively.
Amplitude expressions not shown here and formulas for
differential decay widths are provided in Appendices~\ref{app1} and 
\ref{app2}, respectively.

\begin{figure*}[t]
\includegraphics[width=1.0\textwidth]{comp-data2}
 \caption{\label{fig:comp-data2}
Invariant mass ($M_f$) distributions
for $e^+e^-\to \gamma f$ where final states $f$ are indicated in each panel.
The data are from Ref.~\cite{x3872_bes3_jpsipipi} in (a);
\cite{x3872_bes3_jpsi-omega} in (b);
\cite{x3872_bes3_DstarDbar} in (c).
}
 \end{figure*}

\section{Results}
By fitting the experimental distributions\footnote{\label{hel}
For a decay chain $B\to X c$ and $X\to ab$,
a helicity angle [Fig.~\ref{fig:comp-data}(f)] is
given with momenta in the $X$-at-rest frame as
$\cos\theta_{X}\equiv {\bm{q}_a\cdot\bm{q}_{B}\over|\bm{q}_a||\bm{q}_{B}|}$;
$a=D^{*0} (\bar{D}^{*0})$ for Figs.~\ref{fig:comp-data}(f) and \ref{fig:angle}(b),
and $a=J/\psi$ or $\psi'$ for Figs.~\ref{fig:angle}(a),
\ref{fig:angle}(c), and \ref{fig:angle}(d).}
in Figs.~\ref{fig:comp-data}--\ref{fig:comp-data2} and the ratios in
Tables~\ref{tab:RB}--\ref{tab:RB2},
we obtain models A (default) and B that include all mechanisms
of Fig.~\ref{fig:diag} (23 fit parameters).
For comparison, we also construct an $\eta_{c2}$-less
($\slashed{\eta}\;\!\!_{c2}$) model by removing the mechanism in
Fig.~\ref{fig:diag}(c) (16 parameters).
The parameter sets and experimental inputs (resolutions, efficiencies,
and kinematic selections) into our calculations are summarized in 
Appendices~\ref{app1} and \ref{app3}, respectively.

Given that the datasets feature vastly different statistical
uncertainties, 
we employ a weighted analysis to obtain balanced fits.
Consequently, we refrain from quoting formal parameter uncertainties, as
they may not fully capture the underlying statistical and systematic effects.
The pole positions are extracted from the models and shown in Table~\ref{tab:pole}.
Within our model variants, these pole locations are stable.
$X(3872)$ and $\eta_{c2}$ poles are slightly below and above the
$D^{*0}\bar{D}^0$ threshold (3871.7~MeV), respectively.
We also find a virtual pole, called $W_{c1}$~\cite{wc1_1,wc1_2}, 
near the $D^{*+}D^-$ threshold ($D^{*+}D^-$ unphysical sheet). 
Our $W_{c1}$ pole locations are 
similar to that reported in Ref.~\cite{wc1_2}.

\begin{table}[b]
\renewcommand{\arraystretch}{1.3}
\caption{\label{tab:pole} Pole locations (MeV) from models A, B, and 
$\slashed{\eta}\;\!\!_{c2}$.
}
\begin{ruledtabular}
\begin{tabular}{cccc}
 & $X(3872)$ & $\eta_{c2}$ & $W_{c1}$\\
A                           &$3871.4-0.022i$  &$3874.2-0.25i$  &$3882.6+1.5i$ \\
B                           &$3871.4-0.022i$  &$3874.2-0.25i$  &$3882.5+1.4i$ \\
$\slashed{\eta}\;\!\!_{c2}$ &$3871.5-0.025i$  &  --            &$3882.2+1.4i$ \\
\end{tabular}
\end{ruledtabular}
\end{table}

All three models give comparable fits to the lineshape data 
in Figs.~\ref{fig:comp-data}--\ref{fig:comp-data2}.
The differences between the default and 
$\slashed{\eta}\,\!_{c2}$ models are within the experimental
uncertainties. 
The slightly concave shape of the default model in 
Fig.~\ref{fig:comp-data}(f) is due to $\eta_{c2}$.
As demonstrated previously~\cite{x3872_lhcb_pipi,mix_para},
we also confirm in Fig.~\ref{fig:comp-data}(b)
that the $M_{\pi^+\pi^-}$ lineshape is formed by the main
$X(3872)\to J/\psi\rho^0$ contribution ('$J/\psi\rho$' in the figure)
and its interference with 
$X(3872)\to J/\psi\omega$ 
followed by the $\omega\to\rho$ mixing. 
In Fig.~\ref{fig:comp-data}(e), the $M_{D^{*0}\bar{D}^0}$ lineshapes
differ between $B^+$ and $B^0$ decays for
$M_{D^{*0}\bar{D}^0}\gtap 3.88$~GeV.
Our models explain this by the fact that 
the $B^+$-decay amplitude features
the color-favored tree mechanisms,
while the $B^0$-decay amplitude involves only
color-suppressed ones.

The BABAR analysis~\cite{x3872_babar_omega} suggested that 
$X(3872)$ could be a $2^{-+}$ state by analyzing the 
$M_{\pi^+\pi^-\pi^0}$ distribution [Fig.~\ref{fig:comp-data}(d)]
where the $\omega$ tail governs the lineshape.
Our default and 
$\slashed{\eta}\,\!_{c2}$
models are consistent with the data after considering
the resolution ($\sigma=6$~MeV). 
Compared to our $\slashed{\eta}\,\!_{c2}$ model, 
the $1^+ X(3872)$ model in Ref.~\cite{x3872_babar_omega}
produces a lineshape shifted to higher $M_{\pi^+\pi^-\pi^0}$.
This is likely because $m_{X(3872)}$ of Ref.~\cite{x3872_babar_omega} is
heavier at 3873.0~MeV, 
thereby including
a larger $\omega$ contribution
from the higher $M_{\pi^+\pi^-\pi^0}$ region.
Thus, these data suggest that any state slightly above the 
$D^{*0}\bar{D}^0$ threshold
cannot decay appreciably into $s$-wave $J/\psi\omega$ ($0^{++}$,
$1^{++}$, $2^{++}$).
They do not, however, exclude a substantial 
$J/\psi\omega$ decay in a higher partial wave. 
As will be discussed later in relation to 
Fig.~\ref{fig:etac2}, 
the current data instead favor such a possibility for 
$\eta_{c2}$.
For $\eta_{c2}\to J/\psi\omega$ in a $p$ wave, the centrifugal barrier
suppresses the high-mass tail and prevents a shift to higher
energies~\cite{x3872_babar_omega}, as realized by the default model
in Fig.~\ref{fig:comp-data}(d).

\begin{table}[b]
\renewcommand{\arraystretch}{1.3}
\caption{\label{tab:RB}
${\cal R}_{B^{0}/B^+}(f)$
from A, B, and $\eta_{c2}$-less ($\slashed{\eta}\;\!\!_{c2}$) models.
$D^{*0}\bar{D}^{0}$ indicates $D^{*0}\bar{D}^{0}+\mathrm{c.c.}$.
Data are from references indicated in the first row. 
}
\begin{ruledtabular}
\begin{tabular}{cccccc}
$f$ & $J/\psi\pi^+\pi^-$\cite{x3872_belle_2} &$J/\psi\omega$\cite{x3872_babar_omega} & $D^{*0}\bar{D}^0$\cite{x3872_belle_DstarD2} & $J/\psi \gamma$\cite{x3872_belle_jpsi-gamma}& $\psi' \gamma$\\
A                           & 0.59 &    0.64&     0.96&      1.0&     0.51\\
B                           & 0.60 &    0.64&     0.93&      1.0&     0.52\\
$\slashed{\eta}\;\!\!_{c2}$ & 0.61 &    0.52&     0.90&     0.52&     0.52\\
Exp. &$0.50\pm 0.15$&$1.0\pm 0.6$& $1.34^{+0.48}_{-0.42}$&$0.70^{+0.47}_{-0.40}$& --\\
\end{tabular}
\end{ruledtabular}
\end{table}

\begin{table*}
\renewcommand{\arraystretch}{1.3}
\caption{\label{tab:RB2}
${\cal R}_x(f,J/\psi\pi^+\pi^-)$ [Eq.~(\ref{eq:RB})] and ${\cal R}_{x}^{\psi\gamma}$.
(Left) $x=B$.
(Right) $x=e^+e^-$.
See Table~\ref{tab:RB} for other features.
}
\begin{ruledtabular}
\begin{tabular}{ccccc|cccc}
$f$& $J/\psi\omega$\cite{x3872_babar_omega} &
	 $D^{*0}\bar{D}^0$\cite{x3872_belle_DstarD2,x3872_lhcb_jpsi-rho} & $J/\psi
	     \gamma$\cite{x3872_belle_jpsi-gamma,x3872_lhcb_jpsi-rho}&
${\cal R}_{B}^{\psi\gamma}$\cite{x3872_lhcb_jpsi-gamma2}
& $J/\psi\omega$\cite{x3872_bes3_jpsi-omega} &
	 $D^{*0}\bar{D}^0$\cite{x3872_bes3_DstarDbar} & $J/\psi
	     \gamma$\cite{x3872_bes3_DstarDbar}&
${\cal R}_{e^+e^-}^{\psi\gamma}$\cite{x3872_bes3_DstarDbar}
\\
A                           & 0.77 & 12.& 0.20 &  1.6 &     1.9 &     13. &    0.85 &    0.24\\
B                           & 0.77 & 12.& 0.19 &  1.6 &     1.9 &     13. &    0.86 &    0.24\\
$\slashed{\eta}\;\!\!_{c2}$ & 0.95 & 13.& 0.27 & 0.62 &    0.98 &     12. &    0.20 &    0.62\\
Exp. &$0.7\pm 0.3$&$12.2^{+3.0}_{-2.6}$& $0.22\pm 0.06$&$1.67\pm 0.25$
&$1.6\pm 0.4$&$11.77\pm 3.09$& $0.79\pm 0.28$&$-0.04\pm 0.28$\\
\end{tabular}
\end{ruledtabular}
\end{table*}

\begin{figure*}
\includegraphics[width=1.0\textwidth]{smearing}
 \caption{\label{fig:smearing}
(a) $M_{J/\psi\omega}$ distribution for $B^+\to K^+J/\psi\omega$.
The red solid and blue dashed curves are from the default model before and after
 smearing with the experimental resolution, respectively.
The inset shows an enlarged view of the low-yield region.
(b) $M_{D^{*0}\bar{D}^0}$ distribution for $B^+\to K^+(D^{*0}\bar{D}^0+{\rm c.c.})$.
The red solid curve and blue dashed histogram are from the default model before and after
averaging with 2-MeV bin width; the resolution-smearing has
 been applied to both. 
(c) Similar to (b) but for $B^0\to K^0(D^{*0}\bar{D}^0+{\rm c.c.})$.
The vertical axes are arbitrary units.
}
 \end{figure*}

The ratios 
${\cal R}_{B^{0}/B^+}(f) \equiv {\Gamma[B^0 \to K^0 X(3872); X(3872)\to f]
\over \Gamma[B^+ \to K^+ X(3872); X(3872)\to f]}$ 
are presented in Table~\ref{tab:RB}.
This ratio should be independent of $f$, provided that
the $X(3872)$ contribution is well isolated.
The central values of the data are rather dependent on $f$, although
the uncertainties are large.
Our results from all models are consistent with the data,
considering the large uncertainties.
The $\slashed{\eta}\,\!_{c2}$ model gives similar 
${\cal R}_{B^{0}/B^+}(f)$ for different $f$,
following the $X(3872)$ dominance.
For the $\slashed{\eta}\,\!_{c2}$ model,
${\cal R}_{B^{0}/B^+}(f)<1$ can be understood from the facts~\cite{x3872_color_favor}
that:
(i) Dominant color-favored $B^+$ ($B^0$) decays in Fig.~\ref{fig:diag}(a) 
produce $\{D^{*0}\bar{D}^0\}$ ($\{D^{*+}D^-\}$);
(ii) $X(3872)$ couples more strongly to
$\{D^{*0}\bar{D}^0\}$ than to $\{D^{*+}D^-\}$.
For the models A and B, 
${\cal R}_{B^{0}/B^+}(f)$ are more dependent on $f$.
This is because both $X(3872)$ and $\eta_{c2}$ contribute and they
have different partial
decay widths for each $f$, and 
the $X(3872)$ production rate relative to $\eta_{c2}$'s
is different between $B^+$ and $B^0$ decays.

The ratios ${\cal R}_x$ [Eq.~(\ref{eq:RB})] in 
Table~\ref{tab:RB2} are 
crucial quantities for determining the validity of the models.
Our models A and B yield similar ${\cal R}_x$ consistent with the data.
However, the $\slashed{\eta}\;\!\!_{c2}$ model significantly 
 fails to reproduce the experimental 
${\cal R}_{e^+e^-}(J/\psi\gamma,J/\psi\pi^+\pi^-)$,
${\cal R}_{B}^{\psi\gamma}$, and
${\cal R}_{e^+e^-}^{\psi\gamma}$.
In particular, 
the $\slashed{\eta}\;\!\!_{c2}$ model yields 
${\cal R}_{B}^{\psi\gamma}\sim {\cal R}_{e^+e^-}^{\psi\gamma}$ since
$X(3872)\to J/\psi\gamma, \psi'\gamma$ 
occur in the same way in $B$ decays and $e^+e^-$ annihilations.
The models A and B give
${\cal R}_{B}^{\psi\gamma}\sim 6.4\times {\cal R}_{e^+e^-}^{\psi\gamma}$
because $\psi'\gamma$ and $J/\psi\gamma$ are mainly from 
$X(3872)$ and $\eta_{c2}$ decays, respectively,
and the production rate of $\eta_{c2}$ relative to that of 
$X(3872)$ is larger in $e^+e^-$ annihilations than in $B$ decays.
The comparisons in Table~\ref{tab:RB2} indicate that the current data
disfavor the $\slashed{\eta}\;\!\!_{c2}$ model, 
and strongly support the hypothesis that 
$X(3872)$ is accompanied by a nearby state.

Recently, BESIII analyzed 
$e^+e^- \to \omega X(3872)$ data~\cite{x3872_bes3_omegaX} and 
extracted the $X(3872)$ branching-fraction ratio 
${\cal B}[X(3872)\to J/\psi\gamma]/{\cal B}[X(3872)\to J/\psi\pi^+\pi^-]=0.38\pm0.20$,
which is only marginally consistent with 
${\cal R}_{e^+e^-}(J/\psi\gamma,J/\psi\pi^+\pi^-)=0.79\pm 0.28$.
Even if they are confirmed to differ in the future,
the two-state hypothesis can naturally
accommodate the data since 
the cross-section ratio of 
$\sigma[e^+e^- \to \omega \eta_{c2}]/\sigma[e^+e^- \to \omega X(3872)]$
should be different from 
$\sigma[e^+e^- \to \gamma \eta_{c2}]/\sigma[e^+e^- \to \gamma X(3872)]$.

Figure~\ref{fig:smearing} helps clarify why the $\eta_{c2}$ signal has not been
identified experimentally so far. In Fig.~\ref{fig:smearing}(a), the unsmeared
$J/\psi\omega$ spectrum exhibits two nearby narrow structures associated
with $X(3872)$ and $\eta_{c2}$, but after smearing with the experimental
resolution of $\sigma=6.7$~MeV, they are merged into a single broad
enhancement. In the $D^{*0}\bar D^0+{\rm c.c.}$ channel, Figs.~\ref{fig:smearing}(b,c)
show that the bin-averaged spectra can further obscure the $\eta_{c2}$
peak: even when the underlying spectrum contains a clear narrow
structure, averaging over the experimental bin width (2~MeV) washes it out and
leaves only a smooth lineshape. 
These observations
indicate that the absence of an experimentally resolved $\eta_{c2}$ peak
does not contradict its existence, but can naturally result from the
combined effects of finite resolution and bin averaging.

Figure~\ref{fig:smearing}(c) also shows a threshold cusp
at the $D^{*+}D^-$ threshold, enhanced by the nearby $W_{c1}$ virtual
pole. Because $B^0$-decay preferentially proceeds through the
color-enhanced $D^{*+}D^-$ production mechanism, this cusp is more
prominent in $B^0$ decay than in $B^+$ decay. 
This isovector $1^{++}$ pole does not play a visible role to resolve the
radiative-decay ratio tension, as seen in the results for
the $\slashed{\eta}\;\!\!_{c2}$ model that involves 
$W_{c1}$.

\begin{figure}
\includegraphics[width=.49\textwidth]{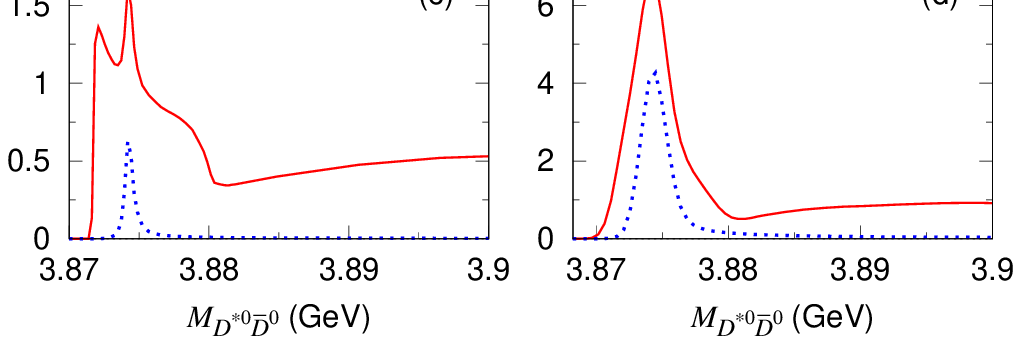}
 \caption{\label{fig:etac2}
(a) [(b)] $M_{J/\psi\omega}$ distribution for $B^+\to K^+J/\psi\omega$ 
[$e^+e^-\to \gamma J/\psi\omega$].
The red solid [blue dashed] curves are from the full calculations
 [$\eta_{c2}$ contributions] of the default model; 
the resolution-smearing has been applied.
(c) [(d)] Similar results for $M_{D^{*0}\bar{D}^0}$ distribution of
$B^+\to K^+(D^{*0}\bar{D}^0+{\rm c.c.})$
[$e^+e^-\to \gamma(D^{*0}\bar{D}^0+{\rm c.c.})$].
The vertical axes are arbitrary units.
}
 \end{figure}

Figure~\ref{fig:etac2} compares the full calculation with the
contribution from only the $\eta_{c2}$-excitation mechanisms of
Fig.~\ref{fig:diag}(c), thereby illustrating the possible size of the
$\eta_{c2}$ component in the relevant observables. 
In the $J/\psi\omega$ and $D^{*0}\bar D^0+{\rm c.c.}$
spectra, the $\eta_{c2}$ contribution is found to be comparable to, and
in some regions even larger than, the $X(3872)$ contribution
in the full result.
At the same time, one
should note an important caveat: the invariant-mass fits in
Figs.~\ref{fig:comp-data} and \ref{fig:comp-data2} alone do not
unambiguously establish the $\eta_{c2}$ contribution, because the
default and $\slashed{\eta}\;\!\!_{c2}$ models provide comparably good descriptions.
Therefore, the $\eta_{c2}$
contributions in Fig.~\ref{fig:etac2} would
have sizable uncertainties. Nevertheless, the existing data still
suggest non-negligible $\eta_{c2}$ effects in the $J/\psi\omega$ and
$D^{*0}\bar D^0+{\rm c.c.}$ channels. For $J/\psi\omega$,
Table~\ref{tab:RB2} indicates that, relative to $J/\psi\pi^+\pi^-$, this
mode is produced more strongly in $e^+e^-$ annihilation than in $B$
decay, and such a trend is naturally explained if the $\eta_{c2}$
contribution is enhanced in $e^+e^-$ reactions as seen in 
Figs.~\ref{fig:etac2}(a) and \ref{fig:etac2}(b).
For $D^{*0}\bar D^0+{\rm c.c.}$, the helicity-angle data in Fig.~\ref{fig:comp-data}(f) tend to
favor a nonzero $\eta_{c2}$ contribution, since the default model gives
a slightly better account of the observed shape.

It is also important to emphasize that $\eta_{c2}$ contributes to the
$J/\psi\pi^+\pi^-$ mode only through the $\rho^0$--$\omega$ mixing effect,
and thus its contribution there is very small. Therefore, the presence
of a nearby $\eta_{c2}$ does not contradict the LHCb determination of $X(3872)$'s
$J^{PC}=1^{++}$ from the $J/\psi\pi^+\pi^-$ channel~\cite{lhcb2013}. 
Rather, the picture
suggested by Fig.~\ref{fig:etac2} is that $\eta_{c2}$ can play an
essential role in channels such as $J/\psi\omega$ and $D^{*0}\bar
D^0+{\rm c.c.}$, while remaining almost invisible in the
$J/\psi\pi^+\pi^-$ mode where the established $X(3872)$ signal
dominates.

Having shown that the two-state hypothesis can account for the existing
data, it is important to test it experimentally.
We therefore provide helicity-angle\footref{hel} distributions for
several processes.
Our predictions for $B^+\to (\psi\gamma)K^+$ and
$e^+e^-\to(\psi\gamma)\gamma$ ($\psi=J/\psi$ or $\psi'$)
are obtained by integrating over $3850 \le M_{\psi\gamma} \le 3890$~MeV
and are shown in Fig.~\ref{fig:angle}(a).
Because this window is wide, we neglect resolution smearing in $M_{\psi\gamma}$.
The angular dependence differs strongly between $J/\psi\gamma$ and
$\psi'\gamma$ because the two channels are dominated by $\eta_{c2}$ and
$X(3872)$ decays, respectively.
If both final states were produced only through $X(3872)$ decays, the
two distributions would instead follow the magenta dashed (green
dash-dotted) curve for $B^+$ decay ($e^+e^-$ annihilation).

\begin{figure}
\includegraphics[width=.49\textwidth]{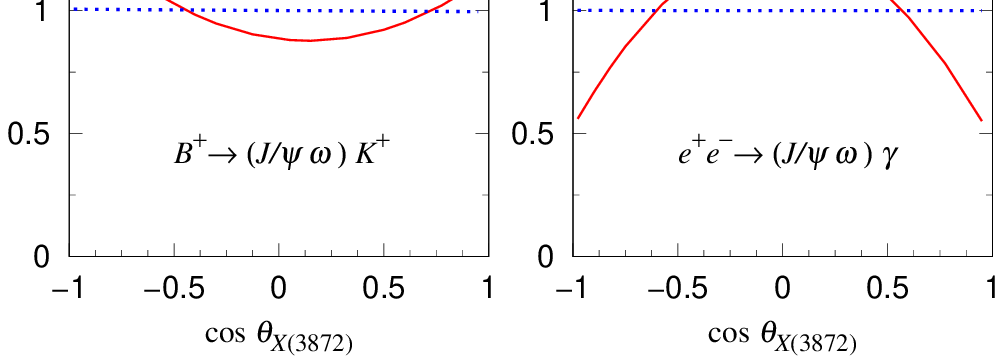}
 \caption{\label{fig:angle}
\red{
Helicity-angle distributions. Panels (a) and (b) show predictions from
the default model, while panels (c) and (d) compare the default and 
$\slashed{\eta}\;\!\!_{c2}$ models.
The vertical axes are arbitrary; however, within each of panels (a) and
(b), the model predicts the relative strengths of different final
states from the same initial state. 
In panels (c) and (d), the curves are normalized to the same area to compare their angular shapes.
}
In (b), $X_L$ and $X_H$ contributions 
correspond to the integrated $M_{D^{*0}\bar{D}^0}$ regions. 
See the main text for details on the resolution smearing and integration procedures.
}
 \end{figure}

We also predict helicity-angle distributions for the
$D^{*0}\bar{D}^0+{\rm c.c.}$ final state [Fig.~\ref{fig:angle}(b)].
Because both $X(3872)$ and $\eta_{c2}$ contribute, we introduce simple
kinematic selections to enhance one component at a time.
We smear the $M_{D^{*0}\bar{D}^0}$ spectrum using the experimental
resolution ($\sigma\sim 1$~MeV) of
Ref.~\cite{x3872_bes3_DstarDbar} and define
$X_L$ and $X_H$ by integrating over $M_{D^{*0}\bar{D}^0}\le 3872$~MeV
and $3872\le M_{D^{*0}\bar{D}^0}\le 3875$~MeV, respectively.
With these definitions, $X_L$ is dominated by $X(3872)$ while $X_H$ is
dominated by $\eta_{c2}$
as seen in Figs.~\ref{fig:etac2}(c) and \ref{fig:etac2}(d), 
leading to distinct angular dependences.
If only a single $X(3872)$ state were present, both $X_L$ and $X_H$
would exhibit a flat distribution.

\red{
Finally, we show helicity-angle distributions for the $J/\psi\omega$ final state
in Figs.~\ref{fig:angle}(c) and \ref{fig:angle}(d), 
comparing the default model with the 
$\slashed{\eta}\;\!\!_{c2}$ model. 
The two models lead to markedly different angular dependences,
reflecting the sizable $\eta_{c2}$ contributions to this channel in the
default model; 
see Figs.~\ref{fig:etac2}(a) and \ref{fig:etac2}(b).
The kinematic cuts used for these distributions are specified in
Table~\ref{tab:exp} in Appendix~\ref{app3}.
}

\section{Summary and outlook}
The radiative-decay ratios reported by LHCb and BESIII in the $X(3872)$
region are difficult to reconcile with a single isolated resonance.
A minimal explanation is a nearby two-state structure: a shallow
$1^{++}$ $D^{*0}\bar{D}^0$ bound state and a $2^{-+}$ charmonium
candidate, $\eta_{c2}$, slightly above the $D^{*0}\bar{D}^0$ threshold.
In this framework, we describe lineshapes and branching ratios
across different production and decay channels,
and we find that an
explicit $\eta_{c2}$ component is required to reproduce the
radiative-ratio tension.
Our helicity-angle predictions provide a direct experimental handle for
testing the two-state scenario and searching for the missing $\eta_{c2}$.

\begin{acknowledgments}
The author acknowledges 
Pablo del Amo S\'anchez,
Christoph Hanhart, 
Atsushi Hosaka,
Yanfeng Li,
Chunhua Li,
Zhiqing Liu,
Kiyoshi Tanida, and 
Junhao Yin
for useful discussions.
This work is in part supported by 
Shandong Province Natural Science Foundation (Grant No. ZR2025MS56).
\end{acknowledgments}

\appendix



\section{Formulas for amplitudes}
\label{app1}

We presented amplitude formulas in Eqs.~(\ref{eq:b_weak}) and (\ref{eq:B_decay}).
The formulas not shown in the main text are given below.
The parameters in these amplitudes are determined by fitting data, as shown in
 Figs.~\ref{fig:comp-data} and \ref{fig:comp-data2} and
 Tables~\ref{tab:RB}--\ref{tab:RB2}.
The parameter values are given in
Tables~\ref{tab:para_B}--\ref{tab:etac2}.


\subsection{Mechanisms of Figs.~\ref{fig:diag}(b) and \ref{fig:diag}(d)}
\label{app1a}

The $Y(4230)$ decay vertices in Fig.~\ref{fig:diag}(b) are 
\begin{eqnarray}
\label{eq:Y_rad}
v^{Y}_{\beta\gamma} &=&
c^{Y}_{\beta\gamma}\,
 f_{\beta}^{0}(q_\beta,\Lambda_Y)
 F_{Y\gamma}^{0}(p_\gamma,\Lambda)\,
(\bm{\epsilon}_{\beta}\times\bm{\epsilon}_{\gamma})
\cdot\bm{\epsilon}_Y,
\end{eqnarray}
with $\beta=\{D^{*0}\bar{D}^0\}$ or $\{D^{*+}D^-\}$.
The $Y\to \gamma\alpha$ amplitude is written, with $E=m_{Y}-E_{\gamma}$, by
\begin{widetext}
\begin{eqnarray}
A^{(b)}_{Y\to \gamma\alpha}
 &=&
(\bm{\epsilon}_{\alpha}\times\bm{\epsilon}_{\gamma})
\cdot\bm{\epsilon}_Y
\sum_{\beta'}
 f_{\alpha}^{0}(q_\alpha,\Lambda_v)\,
G_{\alpha,\beta'}(p_{\gamma},E)
\left(
\sum_{\beta}^{\{D^{*0}\bar{D}^0\},\{D^{*+}D^-\}}
h_{\beta',\beta}\,
c^{Y}_{\beta \gamma}\,
 \sigma^{\Lambda_v\Lambda_Y}_{\beta}(p_{\gamma},E)\right)
 F_{\gamma Y}^{0}(p_\gamma,\Lambda) ,
\label{eq:Y_decay}
\end{eqnarray}
\end{widetext}
where $\bm{\epsilon}_\alpha$ is the total spin polarization of $\alpha$.
For example,
$\bm{\epsilon}_\alpha=\bm{\epsilon}_{D^*}$ or 
$\bm{\epsilon}_{\bar{D}^*}$ for $\alpha=\{D^{*}\bar{D}\}$, and
$\bm{\epsilon}_\alpha=(\bm{\epsilon}_{x}\times \bm{\epsilon}_{J/\psi})/\sqrt{2}$
for $\alpha= J/\psi x$ with $x=\omega, \rho^0, \gamma$.

For tree-level $Y$ decay contributions [Fig.~\ref{fig:diag}(d)],
we use Eq.~(\ref{eq:Y_rad}).


\subsection{Mechanisms of Fig.~\ref{fig:diag}(c)}
\label{app1b}

We use Breit--Wigner forms for 
the $\eta_{c2}$ and $\chi_{c0}(3915)$ excitation amplitudes.
The $\eta_{c2}$-excitation amplitude with
$\alpha=\{D^{*0}\bar{D}^{0}\}$, $J/\psi\omega$, $J/\psi\gamma$, or
$\psi'\gamma$ is given by
\begin{widetext}
\begin{eqnarray}
\label{eq:eta_c2}
A^{\eta_{c2}}_{B\to K\alpha}&=&c^B_{\eta_{c2}K}\,
c^{\eta_{c2}}_\alpha  f_{\alpha}^{1}(q_\alpha,\Lambda_v)\,
{1\over 2 E_{\eta_{c2}}}
{\bm{\epsilon}_\alpha\cdot \bm{p}_K\,
\bm{q}_{\alpha}\cdot \bm{p}_K
-{1\over 3} \bm{\epsilon}_\alpha\cdot \bm{q}_{\alpha}\, p_K^2
\over
E-E_{\eta_{c2}}+
i[\Gamma^{D^*\bar{D}}_{\eta_{c2}}(M_\alpha) +
\Gamma^{\rm nOC}_{\eta_{c2}}]/2}
 F_{K B}^{2}(p_{K},\Lambda)
\ , 
\end{eqnarray}
for $B\to K\alpha$ and $E=m_{B}-E_{K}$, and
\begin{eqnarray}
\label{eq:eta_c2_2}
A^{\eta_{c2}}_{Y\to \gamma\alpha}&=&c^Y_{\eta_{c2}\gamma}\,
c^{\eta_{c2}}_\alpha  f_{\alpha}^{1}(q_\alpha,\Lambda_v)\,
{1\over 2 E_{\eta_{c2}}}
{\bm{\epsilon}_Y\cdot [\bm{p}_{\gamma}\times\left(
{1\over 2} 
\bm{\epsilon}_\alpha\cdot \bm{\epsilon}_\gamma\,
\bm{q}_{\alpha}
+{1\over 2} 
\bm{q}_{\alpha}\cdot \bm{\epsilon}_\gamma\,
\bm{\epsilon}_\alpha
-{1\over 3} 
\bm{q}_{\alpha}\cdot \bm{\epsilon}_\alpha\,
\bm{\epsilon}_\gamma
\right)]
\over
E-E_{\eta_{c2}}
+i[\Gamma^{D^*\bar{D}}_{\eta_{c2}}(M_\alpha) +
\Gamma^{\rm nOC}_{\eta_{c2}}]/2}
 F_{\gamma Y}^{1}(p_\gamma,\Lambda), 
\end{eqnarray}
for $Y\to \gamma\alpha$ and $E=m_{Y}-E_{\gamma}$; $M_\alpha$ is the invariant mass of $\alpha$.
Coupling constants for $B\to K\eta_{c2}$ and 
$Y\to \gamma\eta_{c2}$ are 
$c^B_{\eta_{c2}K}$ and $c^Y_{\eta_{c2}\gamma}$, respectively, and the
$\eta_{c2}\to\alpha$ couplings are 
$c^{\eta_{c2}}_\alpha$.
These coupling constants as well as 
$m_{\eta_{c2}}$ and 
$\Gamma^{\rm nOC}_{\eta_{c2}}$ from non-open-charm (nOC) decay contributions
are fitting parameters.
Since $m_{\eta_{c2}}$ is close to the $D^*\bar{D}$ thresholds,
we consider the energy dependence of 
the $\eta_{c2}$ partial width for $\eta_{c2}\to D^*\bar{D}$ as
\begin{eqnarray}
\Gamma^{D^*\bar{D}}_{\eta_{c2}}(M_\alpha) &=&
2\pi {\left(c^{\eta_{c2}}_{\{D^{*}\bar{D}\}}\right)^2 \over 
M_\alpha^2}
\left[
{q_{D^{*0}\bar{D}^0}^3\over
  (1+q_{D^{*0}\bar{D}^0}^2/\Lambda_v^2)^{5}}
+
{q_{D^{*+}{D}^-}^3\over
  (1+q_{D^{*+}{D}^-}^2/\Lambda_v^2)^{5}}
\right] ,
\end{eqnarray}
where the denominators in the square bracket are from 
the square of the form factor $f_{D^*\bar{D}}^{1}(q_{D^*\bar{D}},\Lambda_v)$;
$M_\alpha = E_{D^*}(q_{D^{*}\bar{D}})+E_{\bar{D}}(q_{D^{*}\bar{D}})$.

Similarly, 
the $\chi_{c0}(3915)$ excitation amplitude with
$\alpha=J/\psi\omega$ is ($\chi_{c0}'\equiv\chi_{c0}(3915)$)
\begin{eqnarray}
\label{eq:chi_c0}
A^{\chi_{c0}'}_{B\to K\alpha}&=&
\bm{\epsilon}_{J/\psi}\cdot\bm{\epsilon}_\omega\,
c^B_{\chi_{c0}'K}\,
c^{\chi_{c0}'}_\alpha  f_{\alpha}^{0}(q_\alpha,\Lambda_v)\,
{1\over 2 E_{\chi_{c0}'}}
{1\over
E-E_{\chi_{c0}'}+
i{\Gamma_{\chi_{c0}'}\over 2}}
 F_{K B}^{0}(p_{K},\Lambda)
\ , 
\end{eqnarray}
for $B\to K\alpha$ and $E=m_{B}-E_{K}$, and
\begin{eqnarray}
\label{eq:chi_c0_2}
A^{\chi_{c0}'}_{Y\to \gamma\alpha}&=&
\bm{\epsilon}_{J/\psi}\cdot\bm{\epsilon}_\omega\,
\bm{\epsilon}_{Y}\cdot\bm{\epsilon}_\gamma\,
c^Y_{\chi_{c0}'\gamma}\,
c^{\chi_{c0}'}_\alpha  f_{\alpha}^{0}(q_\alpha,\Lambda_v)\,
{1\over 2 E_{\chi_{c0}'}}
{1\over
E-E_{\chi_{c0}'}
+i{\Gamma_{\chi_{c0}'}\over 2}}
 F_{\gamma Y}^{0}(p_\gamma,\Lambda),
\end{eqnarray}
\end{widetext}
for $Y\to \gamma\alpha$ and $E=m_{Y}-E_{\gamma}$.
The values of $m_{\chi_{c0}'}$ and $\Gamma_{\chi_{c0}'}$ are taken from Ref.~\cite{x3872_babar_omega}.
In the formulas of this subsection and Appendix~\ref{app1d},
the Lorentz contraction
of the resonance widths is not included; its effect is expected to be
small.

\begin{table*}
\renewcommand{\arraystretch}{1.3}
\caption{\label{tab:para_B}
$B$ decay strength parameters and cutoff for models A, B, and 
$\slashed{\eta}\;\!\!_{c2}$, appearing in formulas indicated in the last column.
Parameters not shown are related to those in the table by
$c^{B^+}_{D^+D^{*-}K^+}=c^{B^+}_{D^{*+}D^-K^+}$,
$c^{B^0}_{D^{*+}D^-K^0}=c^{B^+}_{D^{*0}\bar{D}^0K^+}$,
$c^{B^0}_{D^+D^{*-}K^0}=c^{B^+}_{D^0\bar{D}^{*0}K^+}$,
$c^{B^0}_{D^{*0}\bar{D}^0K^0}=c^{B^+}_{D^{*+}D^-K^+}$, and
$c^{B^0}_{D^0\bar{D}^{*0}K^0}=c^{B^+}_{D^+D^{*-}K^+}$.
}
\begin{ruledtabular}
\begin{tabular}{lcccc}
&\multicolumn{1}{c}{A} & \multicolumn{1}{c}{B} & \multicolumn{1}{c}{$\slashed{\eta}\;\!\!_{c2}$}&Eq.\\\hline 
$c^{B^+}_{D^{*0}\bar{D}^0K^+}$(GeV$^{-1}$)&$  -0.0438            $& $  -0.0115            $& $   0.0505            $&(\ref{eq:b_weak})\\
$c^{B^+}_{D^0\bar{D}^{*0}K^+}$(GeV$^{-1}$)&$   0.0138 +   0.0408i$& $   0.0601 +   0.0097i$& $  -0.0101 -   0.0158i$&(\ref{eq:b_weak})\\
$c^{B^+}_{D^{*+}D^-K^+}$(GeV$^{-1}$)      &$  -0.0103 -   0.0047i$& $   0.0003 +   0.0114i$& $  0.00679 +  0.00668i$&(\ref{eq:b_weak})\\
$c^B_{\eta_{c2}K}$(GeV$^{-1}$)            &   $-$0.118              &   $-$0.0760              &    --    &(\ref{eq:eta_c2})\\
$c^B_{\chi_{c0}'K}$(GeV)                  &\multicolumn{3}{c}{1 (fixed)} &(\ref{eq:chi_c0})\\
$\Lambda_B$(GeV)                          &    0.700	     &      0.700	    &       0.700
&(\ref{eq:b_weak})
\end{tabular}
\end{ruledtabular}
\end{table*}

\begin{table}
\renewcommand{\arraystretch}{1.3}
\caption{\label{tab:para_Y}
$Y(4230)$ decay strength parameters and cutoff.
}
\begin{ruledtabular}
\begin{tabular}{lrrrc}
&\multicolumn{1}{c}{A} & \multicolumn{1}{c}{B} & \multicolumn{1}{c}{$\slashed{\eta}\;\!\!_{c2}$}&Eq.\\\hline 
$c^{Y}_{\{D^{*0}\bar{D}^0\} \gamma}$&    0.0193 &    0.0190 &  $-$0.0237&(\ref{eq:Y_rad})   \\
$c^{Y}_{\{D^{*+}D^-\} \gamma}$      &  $-$0.00384 &  $-$0.00377 & $-$0.00473&(\ref{eq:Y_rad})   \\
$c^Y_{\eta_{c2}\gamma}$             &    $-$0.333 &    $-$0.216 &    --  &(\ref{eq:eta_c2_2})\\
$c^Y_{\chi_{c0}'\gamma}$(GeV)       &\multicolumn{3}{c}{1 (fixed)}&(\ref{eq:chi_c0_2})\\
$\Lambda_Y$(GeV)                    &    1.973 &  1.973 & 1.972&(\ref{eq:Y_rad})   
\end{tabular}
\end{ruledtabular}
\end{table}

\begin{table}
\renewcommand{\arraystretch}{1.3}
\caption{\label{tab:hba}
Coupling constants $h_{\beta,\alpha}$ and a cutoff $\Lambda_v$
for $\{D^{*0}\bar{D}^0\}-\{D^{*+}D^-\}-J/\psi\omega-J/\psi\rho-J/\psi\gamma-\psi'\gamma$
coupled-channel scattering [Eq.~(\ref{eq:cont-ptl})]. 
Parameters not shown are related to those shown in the table by
$h_{\{D^{*+}D^-\},\{D^{*+}D^-\}}=h_{\{D^{*0}\bar{D}^0\},\{D^{*0}\bar{D}^0\}}$, 
$h_{\{D^{*+}D^-\},J/\psi\omega}=h_{\{D^{*0}\bar{D}^0\},J/\psi\omega}$, and 
$h_{\{D^{*+}D^-\},J/\psi\rho}=-h_{\{D^{*0}\bar{D}^0\},J/\psi\rho}$.
The coupling parameters are also converted to the isospin basis, and are
 shown below the horizontal line. 
With an isospin state $\ket{I,I_z}$, a convention of 
$\ket{D^{(*)+}}=-\ket{1/2,1/2}$ is used.
}
\begin{ruledtabular}
\begin{tabular}{lrrr}
&\multicolumn{1}{c}{A} & \multicolumn{1}{c}{B} & \multicolumn{1}{c}{$\slashed{\eta}\;\!\!_{c2}$}\\\hline 
$h_{\{D^{*0}\bar{D}^0\},\{D^{*0}\bar{D}^0\}}$      &   $-$14.9  &     $-$15.0 &    $-$13.7 \\
$h_{\{D^{*0}\bar{D}^0\},\{D^{*+}D^-\}}$            &   $-$6.27  &     $-$6.20 &    $-$4.98 \\
$h_{\{D^{*0}\bar{D}^0\},J/\psi\omega}$             &   0.945  &     0.963 &     1.29 \\
$h_{\{D^{*0}\bar{D}^0\},J/\psi\rho}$               &    1.63  &      1.65 &     1.59 \\
$h_{\{D^{*0}\bar{D}^0\},J/\psi\gamma}$             &     0.00 &  0.000135 &   $-$0.200 \\
$h_{\{D^{*0}\bar{D}^0\},\psi'\gamma}$              &   0.191  &     0.193 &    0.157 \\
$h_{\{D^{*+}D^-\},J/\psi\gamma}$                   &    $-$0.00 &  0.000135 &   $-$0.200 \\
$h_{\{D^{*+}D^-\},\psi'\gamma}$                    &   0.191  &     0.193 &    0.157 \\\hline
$h_{\{D^{*}\bar{D}\}_{I=0},\{D^{*}\bar{D}\}_{I=0}}$&   $-$21.2  &     $-$21.2 &    $-$18.7 \\
$h_{\{D^{*}\bar{D}\}_{I=0},J/\psi\omega}$          &   $-$1.34  &     $-$1.36 &    $-$1.83 \\
$h_{\{D^{*}\bar{D}\}_{I=0},J/\psi\gamma}$          &     0.00 & $-$0.000191 &    0.283 \\
$h_{\{D^{*}\bar{D}\}_{I=0},\psi'\gamma}$           &  $-$0.269  &    $-$0.273 &   $-$0.222 \\\hline
$h_{\{D^{*}\bar{D}\}_{I=1},\{D^{*}\bar{D}\}_{I=1}}$&   $-$8.67  &     $-$8.77 &    $-$8.75 \\
$h_{\{D^{*}\bar{D}\}_{I=1},J/\psi\rho}$            &    2.31  &      2.33 &     2.25 \\
$h_{\{D^{*}\bar{D}\}_{I=1},J/\psi\gamma}$          &        -- &      -- &      -- \\
$h_{\{D^{*}\bar{D}\}_{I=1},\psi'\gamma}$           &        -- &      -- &      -- \\\hline
$\Lambda_v$ (GeV)                                  &     0.950 &    0.952&    1.046
\end{tabular}
\end{ruledtabular}
\end{table}

\begin{table}
\renewcommand{\arraystretch}{1.3}
\caption{\label{tab:etac2}
$\eta_{c2}$ and $\chi_{c0}(3915)$ mass, width, and decay strength
 parameters defined in Eqs.~(\ref{eq:eta_c2})--(\ref{eq:chi_c0_2}).
}
\begin{ruledtabular}
\begin{tabular}{lrrr}
&\multicolumn{1}{c}{A} & \multicolumn{1}{c}{B} & \multicolumn{1}{c}{$\slashed{\eta}\;\!\!_{c2}$}\\\hline 
$m_{\eta_{c2}}$(MeV)                     &    3874.2  & 3874.2  & -- \\
$\Gamma^{\rm nOC}_{\eta_{c2}}$(MeV)      &    0.5 (fixed)&0.5 (fixed)& -- \\
$c^{\eta_{c2}}_{\{D^{*0}\bar{D}^0\}}$    &    $-$0.112&     0.171& -- \\
$c^{\eta_{c2}}_{J/\psi\omega}$           &   $-$0.0731&    $-$0.113& -- \\
$c^{\eta_{c2}}_{J/\psi\gamma}$           &   0.00492&  $-$0.00761& -- \\
$c^{\eta_{c2}}_{\psi'\gamma}$            &      0.00&     $-$0.00& -- \\
$m_{\chi_{c0}'}$(MeV)                    &&3919.9 (fixed)&\\
$\Gamma_{\chi_{c0}'}$(MeV)               &&31.0 (fixed)&\\
$c^{\chi_{c0}'}_{J/\psi\omega}$(MeV)     &&1 (fixed)
\end{tabular}
\end{ruledtabular}
\end{table}


\subsection{$\rho^0-\omega$ mixing}
\label{app1c}

Let us define an amplitude 
$\bar{A}_{B\to K\alpha}$ by
 collecting the amplitudes discussed so far
in Eqs.~(\ref{eq:B_decay}), (\ref{eq:eta_c2}), and (\ref{eq:chi_c0}) as
\begin{eqnarray}
\label{eq:barA}
\bar{A}_{B\to K\alpha}=A^{(a)}_{B^+\to K^+\alpha}+A^{\eta_{c2}}_{B\to K\alpha}
+A^{\chi_{c0}'}_{B\to K\alpha} .
\end{eqnarray}
In addition, for $\alpha=J/\psi\rho^0$ or $J/\psi\omega$, we consider 
the $\rho^0-\omega$ mixing.
Thus, let us define an amplitude $A_{B\to K\alpha}$ by
\begin{eqnarray}
A_{B\to KJ/\psi\rho^0} &=& \bar{A}_{B\to KJ/\psi\rho^0}
-{\epsilon_{\rho\omega}\over
m_\omega^2 - \bar{m}_\omega^2 + i \bar{m}_\omega\bar{\Gamma}_{\omega}} 
\nonumber \\ 
&&
\times \bar{A}_{B\to KJ/\psi\omega} ,
\label{eq:mix1}
\end{eqnarray}
for $\alpha=J/\psi\rho^0$, and 
\begin{eqnarray}
A_{B\to KJ/\psi\omega} &=& \bar{A}_{B\to KJ/\psi\omega}
-{\epsilon_{\rho\omega}\over
m_\rho^2 - \bar{m}_\rho^2 + i \bar{m}_\rho\bar{\Gamma}_{\rho}} 
\nonumber \\ 
&&\times
\bar{A}_{B\to KJ/\psi\rho^0} ,
\label{eq:mix2}
\end{eqnarray}
for $\alpha=J/\psi\omega$, and 
\begin{eqnarray}
A_{B\to K\alpha} &=& \bar{A}_{B\to K\alpha}, 
\label{eq:mix3}
\end{eqnarray}
otherwise.
We have introduced the $\rho^0-\omega$ mixing parameter 
$\epsilon_{\rho\omega}=3.35\times 10^{-3}$~GeV$^2$~\cite{mix_para},
and the nominal vector meson mass ($\bar{m}_V$) and width
($\bar{\Gamma}_V$) for $V=\rho^0$ and $\omega$;
$\bar{m}_{\rho^0}=775$~MeV, $\bar\Gamma_{\rho^0}=150$~MeV,
$\bar{m}_{\omega}=783$~MeV, and
$\bar{\Gamma}_{\omega}=8.49$~MeV~\cite{pdg}.
The mass symbols $m_\omega$ in Eq.~(\ref{eq:mix1}) and 
$m_\rho$ in Eq.~(\ref{eq:mix2}) are treated as variables, as will be
seen below in Appendix~\ref{app2a}.

Similarly, we include the $\rho^0-\omega$ mixing for $Y\to\gamma\alpha$.
We define an amplitude 
$\bar{A}_{Y\to \gamma\alpha}$ by
summing the amplitudes in 
Eqs.~(\ref{eq:Y_decay}), 
(\ref{eq:eta_c2_2}), and (\ref{eq:chi_c0_2}) as
\begin{eqnarray}
\label{eq:barA2}
\bar{A}_{Y\to \gamma\alpha}=
A^{(b)}_{Y\to \gamma\alpha}
+A^{\eta_{c2}}_{Y\to \gamma\alpha}
+A^{\chi_{c0}'}_{Y\to \gamma\alpha},
\end{eqnarray}
and the amplitudes $A_{Y\to \gamma\alpha}$, including the mixing effect,
are obtained from Eqs.~(\ref{eq:mix1})--(\ref{eq:mix3})
by replacing $B\to Y$ and $K\to \gamma$.


\subsection{Widths in Eq.~(\ref{eq:sigma})}
\label{app1d}

For $\alpha=J/\psi\rho^0$,
we use in Eq.~(\ref{eq:sigma})
$\Gamma_{J/\psi}=0$ and
\begin{eqnarray}
\Gamma_{\rho^0}(M)
 = 
\bar\Gamma_{\rho^0}
{k^3\over\bar{k}^3}
\left({\bar{m}_{\rho^0}\over M}\right)^2
{
(1+\bar{k}^2/\Lambda_V^2)^5 \over
(1+{k}^2/\Lambda_V^2)^5
},
\label{eq:rho_wid}
\end{eqnarray}
with $\Lambda_V=1$~GeV,
and the last factor is the ratio of squared form factors.
The argument $M$ denotes the invariant mass of $\pi^+\pi^-$ 
from $\rho^0$ decay, and $k$ is the $\pi^+$ momentum 
in the $\rho^0$-at-rest frame;
$\bar{k}$ is for $M=\bar{m}_{\rho^0}$.
In Eq.~(\ref{eq:sigma}), $M$ is calculated using
\begin{eqnarray}
          M^2=\left[\sqrt{E^2 - p^2} - E_{J/\psi}(q)\right]^2 - q^2 .
\label{eq:M}
\end{eqnarray}

For $\alpha=J/\psi\omega$, 
we again use Eq.~(\ref{eq:M}) to determine $M$, and pass it to the total
$\omega$ width of
 \begin{eqnarray}
\Gamma_{\omega}(M) = 
\Gamma_{\omega\to\pi^0\gamma}(M)
+\Gamma_{\omega\to\pi^+\pi^-\pi^0}(M) , 
\label{eq:omega_wid0}
\end{eqnarray}
where
 \begin{eqnarray}
\Gamma_{\omega\to\pi^0\gamma}(M) = 
\bar{\Gamma}_{\omega}{\cal B}(\omega\to\pi^0\gamma)
\!\left[
{\bar{m}_{\omega} (M^2 - m_{\pi^0}^2)
\over M (\bar{m}_{\omega}^2 - m_{\pi^0}^2)}
\right]^3 \!\!\!\!\!,
\label{eq:omega_wid2}
\end{eqnarray}
with ${\cal B}(\omega\to\pi^0\gamma)=0.0840$~\cite{pdg}, and 
 \begin{eqnarray}
\Gamma_{\omega\to\pi^+\pi^-\pi^0}(M)
= c^\omega_1\left(  {M\over 1{\rm GeV}} + c^\omega_2\right)^{c^\omega_3} ,
\label{eq:omega_wid1}
\end{eqnarray}
for $M\ge 2m_{\pi^\pm}+m_{\pi^0}$, and 
$\Gamma_{\omega\to\pi^+\pi^-\pi^0}(M)=0$ otherwise;
the parameters are
$c^\omega_1=1.061\times 10^{-3}$~MeV,
$c^\omega_2=0.663$,
and 
$c^\omega_3=24.148$.
To determine these parameters,
we consider
a $\omega\to\rho\pi$ ($\rho\to\pi\pi$) model including a 
Breit-Wigner $\rho$ amplitude; all possible $\rho\pi$ charge states are
included coherently.
The $\omega\to\rho\pi$ coupling constant is fixed to reproduce 
$\Gamma_{\omega\to\pi^+\pi^-\pi^0}(\bar{m}_{\omega})=
\bar{\Gamma}_{\omega} {\cal B}(\omega\to\pi^+\pi^-\pi^0)$
with
${\cal B}(\omega\to\pi^+\pi^-\pi^0)=0.892$~\cite{pdg}.
Then, the $M$ dependence of $\Gamma_{\omega\to\pi^+\pi^-\pi^0}$
from this model is fitted with the parameterization of
Eq.~(\ref{eq:omega_wid1})
in the region $M<0.8$~GeV.

For $\alpha=\{D^{*+}{D}^-\}$,
$M$ is determined using Eq.~(\ref{eq:M}) where 
$E_{J/\psi}$ is replaced by $E_{D^-}$, and
\begin{eqnarray}
\Gamma_{D^{*+}}(M)&=&
\Gamma_{D^{*+}\to D^0\pi^+}(M) \, \theta(M-m_{D^0}-m_{\pi^+})
\nonumber\\
&&+\Gamma_{D^{*+}\to D^+\pi^0}(M) \, \theta(M-m_{D^+}-m_{\pi^0})
\nonumber\\
&&+\Gamma_{D^{*+}\to D^+\gamma} \, \theta(M-m_{D^+}), 
\label{eq:Dstar_wid0}
\end{eqnarray}
where $\theta(x)$ is the Heaviside step function, and 
\begin{eqnarray}
\Gamma_{D^{*+}\to D^0\pi^+}(M)
 &=& 
{2\over 3}2\pi {g_{D^*}^2 k_{\pi^+}^3\over M^2} 
\left(1+{k_{\pi^+}^2\over\Lambda_V^2}\right)^{\!\!\!-5}\!\!\!\!\!,
\label{eq:Dstar_wid1}
\\
\Gamma_{D^{*+}\to D^+\pi^0}(M)
 &=& 
{1\over 3}2\pi {g_{D^*}^2 k_{\pi^0}^3\over M^2} 
\left(1+{k_{\pi^0}^2\over\Lambda_V^2}\right)^{\!\!\!-5}\!\!\!,
\label{eq:Dstar_wid2}
\end{eqnarray}
and 
$\Gamma_{D^{*+}\to D^+\gamma}=1.33$~keV;
$g_{D^*}=0.948$;
$\Gamma_{D^{*+}}(m_{D^{*+}})=83.4$~keV.

Similarly, for $\alpha=\{D^{*0}\bar{D}^{0}\}$, we use
\begin{eqnarray}
\Gamma_{D^{*0}}(M)&=&
\Gamma_{D^{*0}\to D^0\pi^0}(M) \, \theta(M-m_{D^0}-m_{\pi^0})
\nonumber\\
&&+\Gamma_{D^{*0}\to D^+\pi^-}(M) \, \theta(M-m_{D^+}-m_{\pi^-})
\nonumber\\
&&+\Gamma_{D^{*0}\to D^0\gamma} \, \theta(M-m_{D^0}), 
\label{eq:Dstar0_wid0}
\end{eqnarray}
with
\begin{eqnarray}
\Gamma_{D^{*0}\to D^0\pi^0}(M)
 &=& 
{1\over 3}2\pi {g_{D^*}^2 k_{\pi^0}^3\over M^2} 
\left(1+{k_{\pi^0}^2\over\Lambda_V^2}\right)^{\!\!\!-5}\!\!\!,
\label{eq:Dstar0_wid1}
\\
\Gamma_{D^{*0}\to D^+\pi^-}(M)
 &=& 
{2\over 3}2\pi {g_{D^*}^2 k_{\pi^-}^3\over M^2} 
\left(1+{k_{\pi^-}^2\over\Lambda_V^2}\right)^{\!\!\!-5}\!\!\!\!\!,
\label{eq:Dstar0_wid2}
\end{eqnarray}
and $\Gamma_{D^{*0}\to D^0\gamma}=19.4$~keV;
$\Gamma_{D^{*0}}(m_{D^{*0}})=55$~keV.
\\


\section{Formulas for differential decay width}
\label{app2}

The differential decay width (Dalitz plot distribution) for 
$B\to Kf_1f_2$ ($f_1f_2$ form a final state $f$) is given by
\begin{eqnarray}
\frac{d^2\Gamma_{B\to Kf}}{dM_{f_1f_2}^2 dM_{Kf_1}^2} &=&
\frac{1}{(2\pi)^3}\frac{1}{32m_B^3} \frac{1}{2s_{B}+1} 
\nonumber\\
&&\times
\sum_{\rm spins}
|{\cal M}_{B\to Kf}|^2,
\label{eq:dalitz-unpol}
\end{eqnarray}
where the initial spin states are averaged, and 
all the final spin states are summed.
The invariant amplitude ${\cal M}_{B\to Kf}$ is related to 
$A_{B\to Kf}$ by
\begin{eqnarray}
{\cal M}_{B\to Kf} &=& - (2\pi)^3
\sqrt{2m_B} \sqrt{2E_K} \sqrt{2E_{f_1}} \sqrt{2E_{f_2}} 
\nonumber\\
&& \times A_{B\to Kf}, 
\label{eq:M_BKf}
\end{eqnarray}
where
\begin{eqnarray}
\label{eq:Abkf1}
 A_{B\to Kf} = {1\over\sqrt{2}} A_{B\to K\alpha} + v^{B}_{fK}
\end{eqnarray}
for $f=D^{*0}\bar{D}^0$ or $D^{0}\bar{D}^{*0}$, and 
$\alpha=\{D^{*0}\bar{D}^0\}$, and 
\begin{eqnarray}
\label{eq:Abkf2}
 A_{B\to Kf} = A_{B\to K\alpha} , 
\end{eqnarray}
otherwise; we do not consider 
$f=D^{*+}{D}^-$ and $D^{+}{D}^{*-}$ in this work.
The amplitude
$A_{B\to K\alpha}$ has been defined in 
Eqs.~(\ref{eq:barA})--(\ref{eq:mix3});
$v^{B}_{fK}$ is defined in Eq.~(\ref{eq:b_weak});
$A_{B^+\to K^+D^{*0}\bar{D}^0}\ne A_{B^+\to K^+D^{0}\bar{D}^{*0}}$
since 
$v^{B^+}_{D^{*0}\bar{D}^0K^+}\ne v^{B^+}_{D^{0}\bar{D}^{*0}K^+}$, while 
$A_{B^0\to K^0D^{*0}\bar{D}^0}=A_{B^0\to K^0D^{0}\bar{D}^{*0}}$
since 
$v^{B^0}_{D^{*0}\bar{D}^0K^0}=v^{B^0}_{D^{0}\bar{D}^{*0}K^0}$.

Regarding $e^+e^-\to \gamma f_1f_2$,
we first obtain ${\cal M}_{Y\to \gamma f}$ from 
Eqs.~(\ref{eq:M_BKf})--(\ref{eq:Abkf2})
by replacements $B\to Y$, $K\to \gamma$, and, then,
$v^{Y}_{f\gamma}\to {1\over\sqrt{2}}v^{Y}_{\alpha\gamma}$
in Eq.~(\ref{eq:Abkf1}).
$A_{Y\to \gamma\alpha}$ is
defined in Eq.~(\ref{eq:barA2}) and the following sentence;
$v^{Y}_{\alpha\gamma}$ is defined in Eq.~(\ref{eq:Y_rad}).
Then we calculate 
differential cross sections for
$e^+e^-\to \gamma f_1f_2$
using Eq.~(23) of Ref.~\cite{ee-cc} with
$\bar{\cal M}^\mu_{abc} = 
{\cal M}_{Y\to \gamma f}/[2 M_Y (E-M_Y+i{\Gamma_Y\over 2})]$; the
superscript $\mu$ denotes the polarization of $Y(4230)$.

For $B\to KD^{*0}\bar{D}^0$,
we also calculate the helicity-angle ($\theta_{X(3872)}$) 
distribution; $\theta_{X(3872)}$ is 
the angle between the $D^{*0}$ and $B$ momenta 
in the $D^{*0}\bar{D}^0$ CM frame.
We use the formula ($f=D^{*0}\bar{D}^0$ or $D^{0}\bar{D}^{*0}$):
\begin{eqnarray}
\frac{d\Gamma_{B\to Kf}}{d \cos\theta_{X(3872)}}
\! &=& \!
\int\!
\frac{d^2\Gamma_{B\to Kf}}{d
\cos\theta_{X(3872)}dM_{f_1f_2}}
dM_{f_1f_2},
\end{eqnarray}
with
\begin{eqnarray}
\frac{d^2\Gamma_{B\to Kf}}{d
\cos\theta_{X(3872)}dM_{f_1f_2}}
&=&
\frac{d^2\Gamma_{B\to Kf}}{dM_{f_1f_2}^2 dM_{Kf_1}^2}
\nonumber\\
&&\times 4m_B\, p_K\, q_{f_1} ,
\label{eq:helicty}
\end{eqnarray}
where $p_K$ is the $K$ momentum in the total CM frame 
and $q_{f_1}$ is the $f_1$ momentum in the $f_1f_2$ CM frame.


\subsection{$\rho^0\to\pi^+\pi^-$ and $\omega\to\pi^+\pi^-\pi^0$}
\label{app2a}

For describing $B\to K J/\psi \rho^0 (\rho^0\to\pi^+\pi^-)$ 
and $B\to K J/\psi \omega (\omega\to\pi^+\pi^-\pi^0)$, 
$d\Gamma_{B\to K J/\psi V}/dM_{J/\psi V}$ ($V=\rho^0,\omega$)
of Eq.~(\ref{eq:dalitz-unpol})
is extended as follows:
\begin{eqnarray}
 { d^2\Gamma_{B\to K J/\psi\pi_V}
  \over dM_{J/\psi\pi_V}  dM_{\pi_V}}
  &&=
{1\over 2 \pi}
  \left.
\frac{d\Gamma_{B\to K J/\psi V}}{dM_{J/\psi V}}
\right|_{m_{V}=M_{\pi_V}}
\nonumber\\
 \times&&  
 { [M_{\pi_V} / E_{V}]^2 \,
 \Gamma_{V\to\pi_V}(M_{\pi_V})
  \over
|E-E_{J/\psi}-E_{V}+ {i\over 2} \Gamma_{V}(M_{\pi_V})|^2
}
\, ,\nonumber\\
\label{eq:V-pi}
\end{eqnarray}
where $(V,\pi_V)$ is either 
$(\rho^0,\pi^+\pi^-)$ or
$(\omega,\pi^+\pi^-\pi^0)$.
$(d\Gamma_{B\to K J/\psi V}/dM_{J/\psi V})|_{m_{V}=M_{\pi_V}}$
is calculated with Eq.~(\ref{eq:dalitz-unpol}) where 
the phase-space and the momenta involved depend on
$m_{V}(=M_{\pi_V})$, and
the $V$ propagators in Eqs.~(\ref{eq:mix1}) and (\ref{eq:mix2}) are also
functions of $m_{V}$;
otherwise, ${\cal M}_{B\to Kf}$ in Eq.~(\ref{eq:dalitz-unpol})
does not depend on $m_{V}$.
The nominal masses for $\rho^0$ 
and $\omega$
enter Eq.~(\ref{eq:V-pi}) through
$E_{V}=\sqrt{\bar{m}^2_{V}+\bm{p}^2_{V}}$.
The total and  partial $V$ decay widths are denoted by
$\Gamma_{V}$ and $\Gamma_{V\to\pi_V}$, respectively,
and the formulas are given in Eqs.~(\ref{eq:rho_wid}) and 
(\ref{eq:omega_wid0})--(\ref{eq:omega_wid1});
$\Gamma_{\rho^0\to\pi^+\pi^-}=\Gamma_{\rho^0}$.

For $e^+e^-\to \gamma J/\psi \rho^0 (\rho^0\to\pi^+\pi^-)$ 
and $e^+e^-\to \gamma J/\psi \omega (\omega\to\pi^+\pi^-\pi^0)$,
we extend the cross-section formula 
in Eq.~(23) of Ref.~\cite{ee-cc} 
in a manner similar to Eq.~(\ref{eq:V-pi}).
\\


\section{Experimental information}
\label{app3}

Experimental information considered in our calculations is summarized in
Table~\ref{tab:exp}.
The resolution models used for the 
$M_{J/\psi\pi^+\pi^-}$ and $M_{D^{*0}\bar{D}^0}$
distributions, which are referenced in Table~\ref{tab:exp},
are described in Appendix~\ref{app4}.
In the table, we also refer to the following formulas:
\begin{eqnarray}
\label{eq:resol-pipi}
\sigma(M_{\pi^+\pi^-}) &=& 
2.39 (1-\exp(- M_{\pi^+\pi^-}/220.3)) \nonumber\\
&&-5.4\exp(-M_{\pi^+\pi^-}/ 220.3) ,\\
\epsilon(M_{\pi^+\pi^-}) &=& 
0.966+1.345\times 10^{-3} (M_{\pi^+\pi^-}-700)\nonumber\\
&&+1.607\times 10^{-6} (M_{\pi^+\pi^-}-700)^2,
\label{eq:eff-pipi}
\end{eqnarray}
with $\sigma(M_{\pi^+\pi^-})$ and 
$M_{\pi^+\pi^-}$ being in MeV.
%


\section{Resolution models for $M_{J/\psi\pi^+\pi^-}$ and $M_{D^{*0}\bar{D}^0}$ distributions}
\label{app4}

To compare theoretical lineshapes with invariant-mass
distribution data,
we smear the theory input $F_{\rm th}(x')$ with an effective detector-resolution
function $R(x,x')$ as
\begin{equation}
  F_{\rm sm}(x)
  =
  \int dx'\,R(\Delta)\,F_{\rm th}(x'),
  \label{eq:conv_appendix}
\end{equation}
with $\Delta \equiv x-x'$.
Here $x$ is the observed invariant mass and $x'$ is the true invariant
mass.


\subsection{Crystal Ball function}
\label{app4a}

We use a one-sided Crystal Ball (CB) function~\cite{Skwarnicki:1986xj}
together with a Gaussian. 
The lower-side-tail CB function is
\begin{equation}
  {\rm CB}_{-}(\Delta;\sigma,\alpha,n)
  =
  {\cal N}_{\rm CB}
  \begin{cases}
    \exp\!\left(-\dfrac{t^2}{2}\right), & t> -|\alpha|, \\[1.0ex]
    A\,(B-t)^{-n}, & t\le -|\alpha|,
  \end{cases}
  \label{eq:CBminus_appendix}
\end{equation}
with
\begin{equation}
  t \equiv \frac{\Delta}{\sigma},
  \quad
  A=\left(\frac{n}{|\alpha|}\right)^n
    \exp\!\left(-\frac{|\alpha|^2}{2}\right),
  \quad
  B=\frac{n}{|\alpha|}-|\alpha|.
  \label{eq:CBminus_pars_appendix}
\end{equation}
Here ${\cal N}_{\rm CB}$ is fixed by
\begin{equation}
  \int d\Delta\,{\rm CB}_{-}(\Delta;\sigma,\alpha,n)=1.
\end{equation}
In this convention the tail lies at $\Delta<0$.
The reversed and upper-side-tail form is given by
\begin{equation}
  {\rm CB}_{+}(\Delta;\sigma,\alpha,n)
  =
  {\rm CB}_{-}(-\Delta;\sigma,\alpha,n). 
  \label{eq:CBplus_appendix}
\end{equation}


\subsection{$M_{J/\psi\pi^+\pi^-}$ distribution}
\label{app4b}

For fitting the LHCb $J/\psi\pi^+\pi^-$ spectrum~\cite{x3872_lhcb_lineshape},
we use a Crystal-Ball-plus-Gaussian (CBG) resolution model:
\begin{eqnarray}
  R_{\rm LHCb}(\Delta)
  &=&
c\, G(\Delta,\sigma_{\rm gauss}) 
\nonumber\\
&&+ (1-c){\rm CB}_{-}(\Delta;\sigma_{\rm CB},\alpha,n),
  \label{eq:R_LHCb_appendix}
\end{eqnarray}
where
\begin{equation}
  G(\Delta,\sigma)
  =
  \frac{1}{\sqrt{2\pi}\,\sigma}
  \exp\!\left[-\frac{\Delta^2}{2\sigma^2}\right].
  \label{eq:Gaussian_appendix}
\end{equation}
In the actual LHCb analysis, the detector response is modeled by a narrow
Crystal Ball function combined with a wider Gaussian
contribution~\cite{x3872_lhcb_lineshape}. The publication does not
specify the tail orientation explicitly. In the present work, we use the
low-side tail in $\Delta$, i.e.\ Eq.~\eqref{eq:R_LHCb_appendix}, since
this choice is more commonly used.

\begin{table}[t]
  \caption{Resolution-model parameters used in the present analysis.}
  \label{tab:resolution_parameters}
  \begin{ruledtabular}
  \begin{tabular}{lrr}
        & LHCb &Belle \\ \hline
     $\sigma_{\rm gauss}$ (MeV) & 6.0 & 0.059\\
     $\sigma_{\rm CB}$ (MeV) & $2.6$ & 0.10\\
     $\alpha$ & $1.5$ & $-$1.04\\
     $n$ & $3.0$ & $127.18$\\
     $c$ &0.15 & $0.7003$ \\
     $p_0$ (MeV$^{-1}$)& -- & $36.0656$ \\
  \end{tabular}
  \end{ruledtabular}
\end{table}


\subsection{$M_{D^{*0}\bar{D}^0}$ distribution}
\label{app4c}

For fitting the Belle data on 
the $M_{D^{*0}\bar{D}^0}$ spectrum from the 
$D^{*0}\to D^0\pi^0$ mode, 
we use the signal-resolution model described in
Refs.~\cite{x3872_belle_DstarD2,hirata-D}. In this case, the resolution
function is
\begin{eqnarray}
  R_{\rm Belle}(\Delta)
  &=&
  \Bigl[
    c\,G(\Delta,\sigma_{\rm gauss})
    +
    (1-c)\,{\rm CB}_{+}(\Delta;\sigma_{\rm CB},\alpha,n)
  \Bigr] 
\nonumber\\
&&\times  f_{\rm turn\mbox{-}on}(M_{D^{*0}\bar{D}^0}),
  \label{eq:R_Belle_appendix}
\end{eqnarray}
where the upper-side-tail Crystal Ball form is used.
Compared with the LHCb resolution model, the Belle resolution function contains an additional
threshold-related correction factor,
\begin{eqnarray}
  f_{\rm turn\mbox{-}on}(M_{D^{*0}\bar{D}^0})
  &=& \frac{1}{2}
  \left[
    1+\operatorname{erf}\!\bigl(p_0 M_{\rm diff}\bigr)
  \right] ,
  \label{eq:Belle_turnon_appendix}
\end{eqnarray}
with 
$M_{\rm diff}\equiv M_{D^{*0}\bar{D}^0}-m_{D^{*0}}-m_{\bar{D}^0}$.


\subsection{Parameter values}
\label{app4d}

The parameter values used in the present analysis are summarized in
Table~\ref{tab:resolution_parameters}. 
For the LHCb case, the exact numerical values of the
Crystal-Ball-plus-Gaussian parameters are not given explicitly in 
Ref.~\cite{x3872_lhcb_lineshape}.
We therefore use representative
working values, chosen within the range commonly employed in
high-energy-physics mass fits.
For the Belle $\pi^0$ mode, the
parameters are from Refs.~\cite{x3872_belle_DstarD2,hirata-D}.

\begin{table*}
\renewcommand{\arraystretch}{1.3}
\caption{\label{tab:exp}
Experimental information considered in our calculations.
Observables are partial widths $\Gamma_p$ and cross sections $\sigma_p$
 for processes $p$.
The helicity angle $\theta_{X(3872)}$ is denoted by $\theta_X$.
We do not apply cuts to kinematical variables other than those shown in
 'kinematical regions'. 
Resolutions are given by the standard deviations ($\sigma$) of the
 gaussian distribution.
Experimental information in the curly brackets are chosen by ourselves,
 while the others are taken from references in the column 'Ref.'.
Hyphens indicate that resolution and/or efficiency are not considered. 
The calculated observables are shown in tables and figures indicated in
 the last column.
}
\begin{ruledtabular}
\begin{tabular}{llccccc}
process ($p$) & observable & kinematical regions (MeV) & $\sigma$ (MeV) &
		 efficiency & Ref. \\\hline
$B^+\to K^+ J/\psi\pi^+\pi^-$ & $\Gamma_p$ & \{$3850\le M_{J/\psi\pi^+\pi^-}\le 3890$\}& -- & -- & --& Tables~\ref{tab:RB},\ref{tab:RB2} \\
& $d\Gamma_p/dM_{J/\psi\pi^+\pi^-}$ & -- & Eq.~(\ref{eq:R_LHCb_appendix}) & -- &
		     \cite{x3872_lhcb_lineshape}& Fig.~\ref{fig:comp-data}(a)\\
& $d\Gamma_p/dM_{\pi^+\pi^-}$ &  \{$3869\le M_{J/\psi\pi^+\pi^-}\le 3873$\}&Eq.~(\ref{eq:resol-pipi}) & Eq.~(\ref{eq:eff-pipi}) & \cite{x3872_lhcb_pipi}& Fig.~\ref{fig:comp-data}(b)\\
$B^0\to K^0 J/\psi\pi^+\pi^-$ & $\Gamma_p$ & \{$3850\le M_{J/\psi\pi^+\pi^-}\le 3890$\}& -- & -- & --&Table~\ref{tab:RB} \\\hline
$B^+\to K^+ J/\psi\omega$ & $\Gamma_p$\footnote{
$\chi_{c0}(3915)$ contribution is not included.
${\cal B}(\omega\to\pi^+\pi^-\pi^0)=89.2\%$~\cite{pdg} is used to
     convert $\pi^+\pi^-\pi^0$ to $\omega$ final state.},
${d\Gamma_p\over d\cos\theta_X}$
& 
$\left\{\, \begin{tabular}{l} 
$740\le M_{\pi^+\pi^-\pi^0}\le 796.5$\\
$\{3850\le M_{J/\psi\omega}\le 3890\}$
\end{tabular} \right.$
& -- & -- & \cite{x3872_babar_omega}& Tables~\ref{tab:RB},\ref{tab:RB2}\\
& $d\Gamma_p/dM_{J/\psi\omega}$ & $740\le M_{\pi^+\pi^-\pi^0}\le 796.5$& 6.7 & -- & \cite{x3872_babar_omega}& Fig.~\ref{fig:comp-data}(c)\\
& $d\Gamma_p/dM_{J/\psi\omega}$ & $769.5\le M_{\pi^+\pi^-\pi^0}\le 796.5$& 6.7 & -- & \cite{x3872_babar_omega}& Fig.~\ref{fig:comp-data}(c)\\
& $d\Gamma_p/dM_{\pi^+\pi^-\pi^0}$ & $3862.5\le M_{J/\psi\pi^+\pi^-\pi^0}\le 3882.5$
& \{6\}\footnote{This resolution is obtained from fitting the $\eta$ peak in
	     Fig.~1 of Ref.~\cite{x3872_babar_omega}.}
 & -- & \cite{x3872_babar_omega}& Fig.~\ref{fig:comp-data}(d)\\
$B^0\to K^0 J/\psi\omega$ & $\Gamma_p$\footnotemark[\numexpr\value{footnote}-3\relax]
 & 
$\left\{\, \begin{tabular}{l} 
$740\le M_{\pi^+\pi^-\pi^0}\le 805.5$\\
$\{3850\le M_{J/\psi\omega}\le 3890\}$
\end{tabular} \right.$
& -- & -- & \cite{x3872_babar_omega}& Table~\ref{tab:RB}\\\hline
\multirow{3}{*}{
\makecell{$B^{+(0)}\to K^{+(0)} D^{*0}\bar{D}^0$\\[.5mm]
 $\qquad +\,K^{+(0)} D^{0}\bar{D}^{*0}$}}
 & $\Gamma_p$\footnote{$\Gamma_p$ from our full calculations are
     multiplied by 42\%($B^+$) and 58\%($B^0$) to compare with data; 
Fig.~4(right) of Ref.~\cite{x3872_belle_DstarD2} shows their analysis
    result where 
the $X(3872)$ contributions are 
$\sim$42\%($B^+$) and $\sim$58\%($B^0$) of the full contribution
for $M_{D^{*0}\bar{D}^0}\le 3900$~MeV.
}
 &\{$M_{D^{*0}\bar{D}^0}\le 3900$\}& -- & -- & -- & Tables~\ref{tab:RB},\ref{tab:RB2}\\
&  $d\Gamma_p/dM_{D^{*0}\bar{D}^0}$ &	 -- & 
Eq.~(\ref{eq:R_Belle_appendix}) &See Ref.
& \cite{x3872_belle_DstarD2,hirata-D}& Fig.~\ref{fig:comp-data}(e)\\
& $d\Gamma_p/d\cos\theta_X$\footnote{$B^+$ and $B^0$ decay
     results are multiplied by their efficiencies, respectively, and then summed.}
&
	 \{$M_{D^{*0}\bar{D}^0}\le 3880$\} & -- & 
$\left\{\begin{tabular}{l} 
$4.3\!\times\! 10^{-4} (B^+)$\\
$0.7\!\times\! 10^{-4} (B^0)$
\end{tabular} \right.$
& \cite{x3872_babar_DstarD}& Fig.~\ref{fig:comp-data}(f)\\
& $d\Gamma^{X_L}_p\!\!/d\!\,\cos\theta_X$
& \{$M_{D^{*0}\bar{D}^0} \le 3872$\} & 
$\sim$1& --
& \cite{x3872_bes3_DstarDbar}& Fig.~\ref{fig:angle}(b)\\
& $d\Gamma^{X_H}_p\!\!/d\!\,\cos\theta_X$
& \{$3872 \le M_{D^{*0}\bar{D}^0} \le 3875$\} & 
$\sim$1& --
& \cite{x3872_bes3_DstarDbar}& Fig.~\ref{fig:angle}(b)\\
\hline 
$B^{+(0)}\to K^{+(0)} J/\psi\gamma$ & $\Gamma_p$ & \{$3850\le M_{J/\psi\gamma}\le 3890$\}& -- & -- & --& Tables~\ref{tab:RB},\ref{tab:RB2}\\
& $d\Gamma_p/d\cos\theta_X$
& \{$3850\le M_{J/\psi\gamma}\le 3890$\} & -- & 
& -- & Fig.~\ref{fig:angle}(a)\\
$B^{+(0)}\to K^{+(0)} \psi'\gamma$ & $\Gamma_p$ & \{$3850\le M_{\psi'\gamma}\le 3890$\}& -- & -- & --& Tables~\ref{tab:RB},\ref{tab:RB2}\\
& $d\Gamma_p/d\cos\theta_X$
& \{$3850\le M_{\psi'\gamma}\le 3890$\} & -- & 
& -- & Fig.~\ref{fig:angle}(a)\\
\hline 
$e^+e^-\to \gamma J/\psi\pi^+\pi^-$ & $\sigma_p$ & \{$3850\le M_{J/\psi\pi^+\pi^-}\le 3890$\}& -- & -- & --& Table~\ref{tab:RB2}\\
& $d\sigma_p/dM_{J/\psi\pi^+\pi^-}$ & -- & 1.14 & -- & \cite{x3872_bes3_jpsipipi}& Fig.~\ref{fig:comp-data2}(a)\\\hline
$e^+e^-\to \gamma J/\psi\omega$ & $\sigma_p$\footnotemark[\numexpr\value{footnote}-3\relax],
${d\sigma_p\over d\cos\theta_X}$
& $\left\{\, \begin{tabular}{l} 
$720\le M_{\pi^+\pi^-\pi^0}\le 810$\\
$\{3850\le M_{J/\psi\omega}\le 3890\}$
\end{tabular} \right.$
& -- & -- &\cite{x3872_bes3_jpsi-omega}& Table~\ref{tab:RB2}\\
& $d\sigma_p/dM_{J/\psi\omega}$ & $720\le M_{\pi^+\pi^-\pi^0}\le 810$& 4.3 & -- & \cite{x3872_bes3_jpsi-omega}& Fig.~\ref{fig:comp-data2}(b)\\\hline
$e^+e^-\to \gamma(D^{*0}\bar{D}^0+{\rm c.c.})$
 & $\sigma_p$\footnote{$\sigma_p$ from our full calculation is
     multiplied by 59\% to compare with data; 
Fig.~2(e) of Ref.~\cite{x3872_bes3_DstarDbar} shows their analysis
     result where the $X(3872)$ contribution  is
$\sim$59\% of the full contribution
for $M_{D^{*0}\bar{D}^0}\le 3900$~MeV.
}
&\{$M_{D^{*0}\bar{D}^0}\le 3900$\}& -- & -- & -- & Table~\ref{tab:RB2}\\
&  $d\sigma_p/dM_{D^{*0}\bar{D}^0}$ &	 -- & 
$\sim$1& --
& \cite{x3872_bes3_DstarDbar}& Fig.~\ref{fig:comp-data2}(c)\\
& $d\sigma^{X_L}_p\!\!/d\!\,\cos\theta_X$
& \{$M_{D^{*0}\bar{D}^0} \le 3872$\} & 
$\sim$1& --
& \cite{x3872_bes3_DstarDbar}& Fig.~\ref{fig:angle}(b)\\
& $d\sigma^{X_H}_p\!\!/d\!\,\cos\theta_X$
& \{$3872 \le M_{D^{*0}\bar{D}^0} \le 3875$\} & 
$\sim$1& --
& \cite{x3872_bes3_DstarDbar}& Fig.~\ref{fig:angle}(b)\\
\hline
$e^+e^- \to \gamma J/\psi\gamma$ & $\sigma_p$ & \{$3850\le M_{J/\psi\gamma}\le 3890$\}& -- & -- & --& Table~\ref{tab:RB2}\\
& $d\sigma_p/d\cos\theta_X$
& \{$3850\le M_{J/\psi\gamma}\le 3890$\} & -- & 
& -- & Fig.~\ref{fig:angle}(a)\\
$e^+e^-\to \gamma \psi'\gamma$ & $\sigma_p$ & \{$3850\le M_{\psi'\gamma}\le 3890$\}& -- & -- & -- &Table~\ref{tab:RB2}\\
& $d\sigma_p/d\cos\theta_X$
& \{$3850\le M_{\psi'\gamma}\le 3890$\} & -- & 
& -- & Fig.~\ref{fig:angle}(a)
\end{tabular}
\end{ruledtabular}
\end{table*}

\clearpage 



\end{document}